\documentclass[twocolumn,amsmath,amssymb,prx,longbibliography,floatfix]{revtex4-2}

\usepackage{graphicx}
\usepackage{dcolumn}
\usepackage{bm}
\usepackage{color}
\usepackage{notes2bib}

\begin{document}



\title{Programmable activation of quantum emitters in high-purity silicon with focused carbon ion beams}

\author{M.~Hollenbach$^{1}$}
\author{N.~Klingner$^{1}$}
\author{P.~Mazarov$^{2}$}
\author{W.~Pilz$^{2}$}
\author{A.~Nadzeyka$^{2}$}
\author{F.~Mayer$^{2}$}
\author{N.~V.~Abrosimov$^{3}$}
\author{L.~Bischoff$^{1}$}
\author{G.~Hlawacek$^{1}$}
\author{M.~Helm$^{1,4}$}
\author{G.~V.~Astakhov$^{1}$}
\email[E-mail:~]{g.astakhov@hzdr.de}

\affiliation{$^1$Helmholtz-Zentrum Dresden-Rossendorf, Institute of Ion Beam Physics and Materials Research, 01328 Dresden, Germany \\
$^2$Raith GmbH, 44263 Dortmund, Germany \\ 
$^3$Leibniz-Institut f\"ur Kristallz\"uchtung (IKZ), 12489 Berlin, Germany \\
$^4$Technical University of Dresden, 01062 Dresden, Germany
 }

\begin{abstract}
Carbon implantation at the nanoscale is highly desired for the engineering of defect-based qubits in a variety of materials, including silicon, diamond, SiC and hBN. However, the lack of focused carbon ion beams does not allow for the full disclosure of their potential for application in quantum technologies. Here, we develop and use a carbon source for focused ion beams for the simultaneous creation of two types of quantum emitters in silicon, the W and G centers. 
Furthermore, we apply a multi-step implantation protocol for the programmable activation of the G centers with sub-100-nm resolution.
This approach provides a route for significant enhancement of the creation yield of single G centers in carbon-free silicon wafers. 
Our experimental demonstration is an important step towards nanoscale engineering of telecom quantum emitters in silicon of high crystalline quality and isotope purity. 
\end{abstract}

 
\date{\today}

\maketitle
Solid-state single-photon emitters are building blocks for photonic quantum technologies \cite{10.1038/nphoton.2016.186, 10.1002/qute.202300423}. 
One promising implementation is based on optically active atomic defects, also known as color centers. They can be created at the desired positions of optoelectronic devices and nanostructures using focused ion beams (FIB) \cite{10.1088/1361-6463/aad0ec, 10.1063/5.0162597}, providing important resources for quantum networks, on-chip information processing and different sensing modalities \cite{10.1038/s41566-018-0232-2e8e}.
However, the lack of certain ion sources for FIB may prevent scalability of this approach for different material systems. 

A particular example is the G center in silicon, which is a carbon related defect consisting of two carbon substitutional and one silicon interstitial atoms \cite{10.1016/0378-4363(83)90256-5, 10.1103/physrevlett.127.196402}. 
Very recently, it has attracted much attention as a single photon source (SPE) in the fiber-telecommunication O-band \cite{10.1364/oe.397377, 10.1038/s41928-020-00499-0}. 
Wafer scale nanofabrication of single G centers with high probability \cite{10.1038/s41467-022-35051-5} and their integration into nanophotonic cavities with a multiple enhancement of the photoluminescence (PL) \cite{10.1038/s41467-023-38559-6, 10.1038/s41467-023-37655-x} can unlock a pathway for exciting applications in quantum cryptography and photonic quantum processing compatible with the silicon-based semiconductor technology \cite{10.1364/oe.397377, 10.1038/s41928-020-00499-0, 10.1038/s41467-022-35051-5, 10.1038/s41467-023-38559-6, 10.1038/s41467-023-37655-x, 10.1103/physrevlett.126.083602, 10.1063/5.0094715, arXiv:2211.09305, 10.1002/adom.202301608, arXiv:2311.08276, 10.1063/5.0130196, arXiv:2302.10230}. 

The most frequently used protocol for the fabrication of G centers is based upon carbon implantation followed by annealing and activation \cite{10.1002/adfm.201103034, 10.1103/physrevb.97.035303}. 
The skipping or modification of the last step results in randomly distributed single G centers \cite{10.1038/s41928-020-00499-0, 10.1364/oe.482311, 10.1063/5.0097407}. 
Using pulsed laser annealing, G centers can be activated locally in controllable way \cite{10.1038/s43246-024-00486-4, 
10.1103/physrevapplied.21.044014, arXiv:2307.05759}, but 
the diffraction limit of the laser spot and heat diffusion impose restrictions on the locality.  
In an alternative approach, the local activation of single G centers with sub-100-nm resolution has been realized using carbon-rich silicon wafer and Si-FIB \cite{10.1038/s41467-022-35051-5}. 
However, this approach is not compatible with high-purity isotopically enriched $\mathrm{^{28}Si}$-wafers possessing low residual carbon concentration \cite{10.1103/physrevb.98.195201}.  
The existence of a carbon source for C-FIB would allow scalable and programmable engineering of single G centers in optoelectonic devices and photonic  nanostructures fabricated from high-purity materials. 

Based on our long-term experience in the development and production of liquid metal alloy ion sources, a first carbon ion emitting source was developed in the past~\cite{10.1116/1.3253471}.
Due to its instability and very short lifetime, this ion source was never used for the creation of color centers. 
Because of the increased interest of the community in such a source, the source formation process has been reconsidered and optimized, and the current procedure is summarized below. 

The formation of a low melting point alloy from carbon and a second element is hindered by the thermodynamic properties of carbon it self and the tendency to form high melting point carbides with many if not all metals.
However, a composition of 20 mol\% carbon and 80 mol\% cerium, forms an eutectic alloy with a theoretical melting point of $661\mathrm{^\circ C}$ at the thermodynamic equilibrium~\cite{10.31399/asm.hb.v03.a0006247}.
This compound has a very low vapor pressure, an additional requirement for a good liquid metal alloy ion source (LMAIS), of around $8 \times 10^{-15}$\,mbar~\cite{10.1179/cmq.1984.23.3.309}. 
To form and melt the alloy, it can be overheated for a short period of time to a vapor pressure of $10^{-6}$\,mbar at around $1167\mathrm{^\circ C}$~\cite{10.1179/cmq.1984.23.3.309}.
Tungsten emitter were electrochemically etched as well as tantalum emitters were mechanically sharpened and could be wetted by the eutectic alloy.

\begin{figure}[tbp]
    \includegraphics[width=0.45\textwidth]{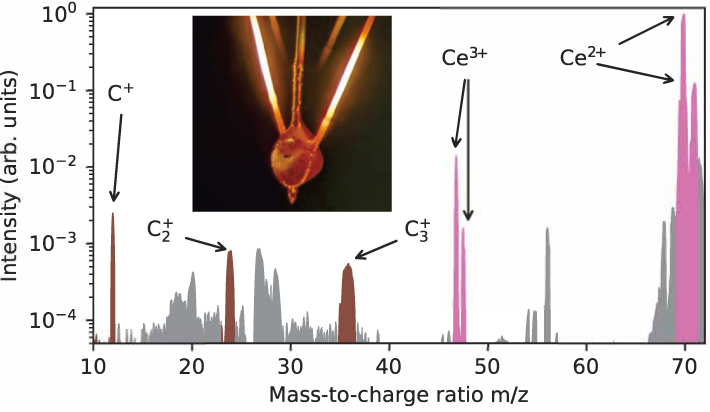}
    \caption{Mass-to-charge ratio spectra and image of CeC emitter. The intensity is normalized to the most abundant beam component of Ce$^{2+}$ ions (86.3\%). Components from atomic carbon ions C$^{+}$ (0.18\%) or carbon cluster ions C$_2^{+}$ (0.06\%) and C$_3^{+}$ (0.04\%) are shown in brown. Ce ions are shown in pink (1.2\% Ce$^{3+}$). Other contributions can be filtered and are indicated in gray. An image of the heated CeC emitter is shown in the inset.} 
\label{fig1}
\end{figure}

The ion beam composition of multiple source has been analyzed by measuring mass-to-charge ratio spectra using a Wien filter (Fig.~\ref{fig1}) in a high resolution test-setup at HZDR as well as in a VELION FIB/SEM at Raith Nanofabrication. 
The most abundant emitted ions are Ce$^{2+}$ ions with 86.3\%, followed by 2.9\% Ce$^{+}$ and heavier cluster ions from Ce and C. 
Even thought 20 mol\% of carbon has been mixed into the alloy, the total intensity of atomic or cluster ions from carbon is only 0.28\%.
Carbon has a much higher first ionization energy of 11.1\,eV compared to 5.5\,eV for Ce, resulting in a higher extraction field required to create carbon ions. 

We measured a contribution of 0.18\% of C$^{+}$ which is still more than sufficient beam current for color center creation. 
An total emission current of $10 \, \mathrm{\mu A}$ was used which, after mass filtering and beam collimation by apertures, resulted in a C$^{+}$ beam current of around 1\,pA.
Additional peaks from impurities found in the complete spectra (not shown) could be identified as Al, Fe, Cu, Lu, Pt, Bi which probably originate from the raw cerium source material.
These contributions can easily be filtered using the Wien filter and do not contribute to the implantation itself.
The observed Al contamination, likely originated from an ionic exchange of Ce and Al from the Al$_2$O$_3$ crucible used in the preparation process.
Consequently tantalum crucibles were used for subsequent melting and source loading runs. 
The C$^+$ ion beam could be focused into a spot size less than 50\,nm, and a total source lifetime of around 80 hours has been achieved.

In order to demonstrate the advantage of our C-FIB, we use a high-purity (100) p-type silicon wafer with resistivity of $3500 \, \mathrm{\Omega \, cm}$ grown by Floating Zone (FZ) technique. 
It has low residual carbon concentration below $10^{15} \, \mathrm{cm^{-3}}$. 
For comparision, we start with the implantation of doubly-charged $\mathrm{Si^{2+}}$ ions at an acceleration potential of $20 \, \mathrm{kV}$, i.e., with an implantation kinetic energy of $40 \, \mathrm{keV}$. 
This energy corresponds to the mean implantation depth of $60 \, \mathrm{nm}$. 
We use a Si-FIB to create a frame with an implantation fluence $\Phi = 1 \times 10^{12} \, \mathrm{cm^{-2}}$. 
After implantation, we measure PL in a confocal microscope at a temperature $T = 7 \, \mathrm{K}$ using a continuous wave excitation at $637 \, \mathrm{nm}$. 
The details of the experimental setup are presented elsewhere \cite{10.1038/s41467-022-35051-5}. 
 
\begin{figure}[t]
\includegraphics[width=.45\textwidth]{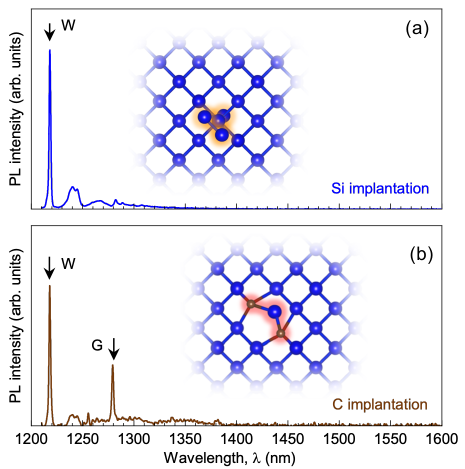}
\caption{PL spectra at $T = 7 \, \mathrm{K}$ form a high-purity silicon wafer after ion implantation. 
(a) Implantation with double-charged $\mathrm{Si^{2+}}$ ions to a fluence of $1 \times 10^{12} \, \mathrm{cm^{-2}}$ at an acceleration potential of $20 \, \mathrm{kV}$. 
Inset shows the atomic configuration of the W center. 
(b) Implantation with single-charged $\mathrm{C^{+}}$ ions to a fluence of $1 \times 10^{12} \, \mathrm{cm^{-2}}$ at an acceleration potential of $40 \, \mathrm{kV}$. 
The inset shows the atomic configuration of the G center.} \label{fig2}
\end{figure}

A PL spectrum after Si-FIB implantation into high-purity silicon is presented in Fig.~\ref{fig2}(a). 
It is dominated by the zero-phonon line (ZPL) at $\lambda_{\mathrm{W}} = 1218 \, \mathrm{nm}$, which is the fingerprint of the W center \cite{10.1364/oe.386450}. 
It consists of three Si interstitials $\mathrm{Si_{i} - Si_i - Si_{i}}$ \cite{10.1021/acsphotonics.2c00336}, as schematically shown in the inset of Fig.~\ref{fig2}(a). 
Obviously, the probability to create G centers, consisting of two C substitutional and one Si interstitial \cite{10.1103/physrevlett.127.196402}, in a high-purity silicon wafer is very low. 
Its atomic configuration $\mathrm{C_{Si} - Si_i - C_{Si}}$ is schematically shown in the inset of Fig.~\ref{fig2}(b). 
The spectroscopic fingerprint of the G center is the ZPL at $\lambda_{\mathrm{G}} = 1278 \, \mathrm{nm}$ \cite{10.1103/physrevb.97.035303}. 
From the comparison of the relative ZPL intensities in Fig.~\ref{fig2}(a), we estimate the probability to create the G centers under Si implantation due to the residual carbon doping $I_{\mathrm{G}} / I_{\mathrm{W}} < 0.05$ (Table~\ref{Probability_GW}). 

 \begin{table}[b]
\caption{Ratio between the ZPL of the G ans W centers depending on the implantation protocol. }
\begin{center}
\begin{tabular}{|c|c|c|}
Single step & Single step & Multi-step \\
$\mathrm{Si^{2+}}$ & $\mathrm{C^{+}}$ & $\mathrm{C^{+}}$ \\
\hline
$< 0.05$ & $>0.4$ & $>7$ \\
\end{tabular}
\end{center}
\label{default}
\label{Probability_GW}
\end{table} 

To increase the probability of the creation of the G centers, we use now our C-FIB with singly-charged $\mathrm{C^{+}}$ ions. 
The implantation kinetic energy of $40 \, \mathrm{keV}$ corresponds to the implantation depth of $125 \, \mathrm{nm}$. 
A typical PL spectrum after implantation fluence $\Phi = 1 \times 10^{12} \, \mathrm{cm^{-2}}$ is shown in Fig.~\ref{fig2}(b). 
The ZPL ratio $I_{\mathrm{G}} / I_{\mathrm{W}} >0.4$ indicates similar density of the W and G centers. 
In contrast to Si implantation, C ions not only displace Si atoms from the lattice, creating interstitials $\mathrm{Si_i}$, but also form carbon substitutional pairs $\mathrm{C_{Si} - C_{Si}}$. 
The dynamical formation of such carbon pairs could be strongly favored under certain implantation conditions \cite{10.1038/s41467-023-36090-2, arXiv:2302.05814}. 

\begin{figure}[t]
\includegraphics[width=.47\textwidth]{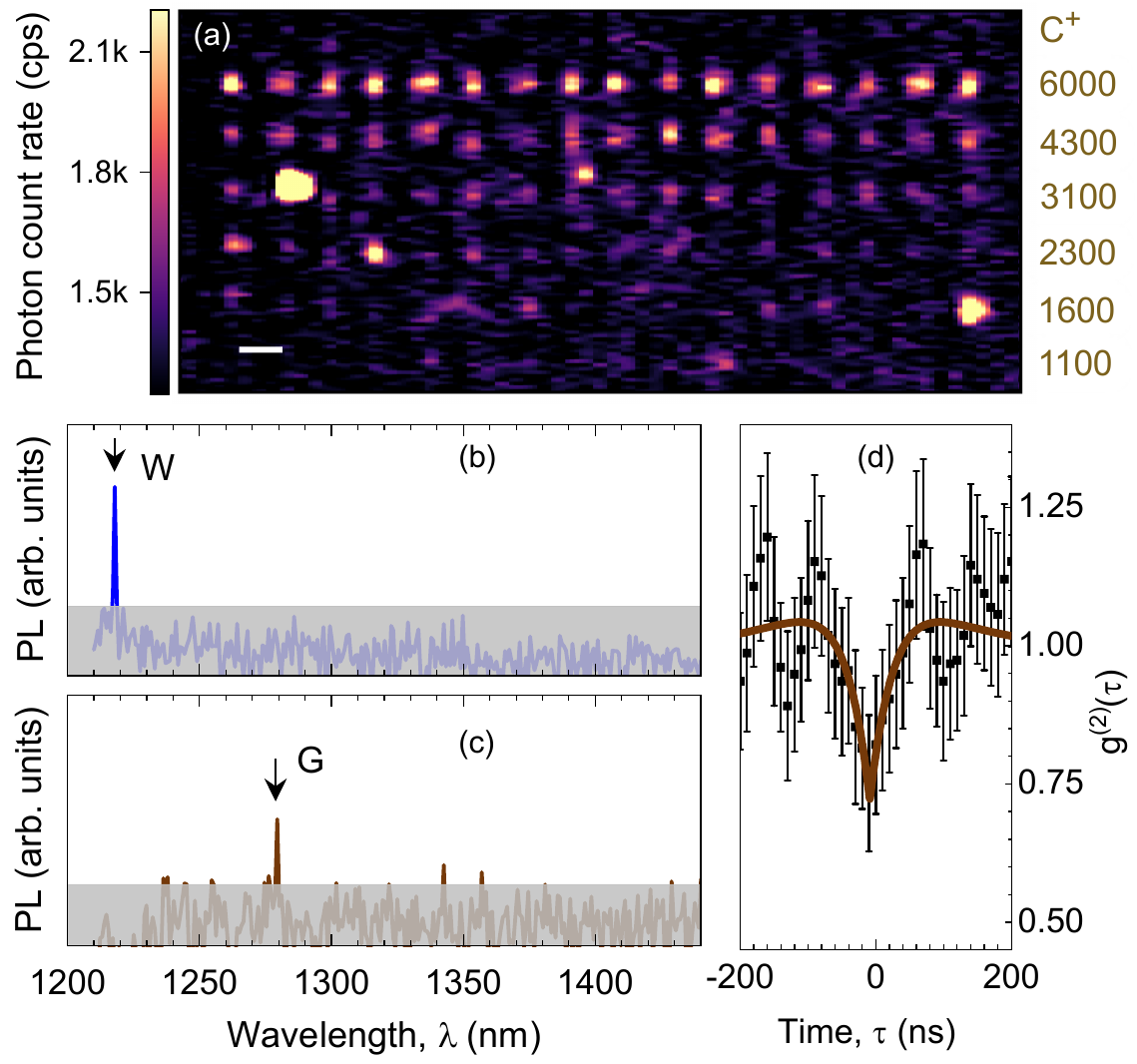}
\caption{Co-creation of W and G centers with implantation of $\mathrm{C^{+}}$ ions. 
(a) Confocal map of the PL below the silicon bandgap after local implantation with a C-FIB. 
The nominal number of implanted $\mathrm{C^{+}}$ ($\bar{n}_{\mathrm{C}}$) ions is indicated for each row. 
The scale bar is $10 \, \mathrm{\mu m}$. 
(b) PL spectrum from one implantation spot in the row with $\bar{n}_{\mathrm{C}} = 2300$.
The arrow indicates the ZPL from the W center. 
The shadowed area represents the noise floor. 
(c) PL spectrum from another implantation spot in the same row. 
The arrow indicates the ZPL from the G center. 
The shadowed area represents the noise floor. 
(d) Second-order autocorrelation function $g^{2} (\tau)$ from one of the implantation spots with a spectrally filtered contribution from the G center. 
The brown solid curve is a fit. 
The error bars represent standard deviation. } \label{fig3}
\end{figure}

To demonstrate the co-creation of the W and G centers, we perform local implantation with C-FIB and create a pattern, similar to our earlier experiments with Si-FiB \cite{10.1038/s41467-022-35051-5}. 
The number of implanted of $\mathrm{C^{+}}$ ions per spot varies from $\bar{n}_{\mathrm{C}} = 1100$ to $\bar{n}_{\mathrm{C}} = 6000$ and the corresponding confocal PL map after implantation is presented in Fig.~\ref{fig3}(a). 
The number of implanted $\mathrm{C^{+}}$ ions is the nominally the same for each row, though the real number statistically fluctuates around $\bar{n}_{\mathrm{C}}$. 
We use a 1200-nm longpass filter to suppress the PL from the silicon bandgap but detect the PL from the W and G centers, including their ZPLs and phonon side bands (PSBs). 

We do not detect any PL from the implanted spots for $\bar{n}_{\mathrm{C}} < 1600$, indicating that this is the threshold value for the creation of single defects. 
Figure~\ref{fig3}(b) shows a PL spectrum from one of the implantation spots with the ZPL $\lambda_{\mathrm{W}} = 1218 \, \mathrm{nm}$, which points at a single W center. 
The PL from another implantation spot is presented in Fig.~\ref{fig3}(c). 
It has the ZPL at $\lambda_{\mathrm{G}} = 1278 \, \mathrm{nm}$, corresponding to the G center. 
We note that the creation of a particular type of defects depends on both local environment and stochastic ion propagation in the crystal. 
To prove the creation of single defects, we use an additional bandpass filter tuned to the ZPL of the G center and measure the second-order autocorrelation function $g^{2} (\tau)$, as shown in Fig.~\ref{fig3}(d). 
From a standard fit of the experimental data \cite{10.1038/ncomms8578}, we find $g^{2}_{corr} (0) = 0.7 \pm 0.1$ and after the background correction $g^2 (0) \rightarrow 0 $. 

\begin{figure}[tbp]
\includegraphics[width=.47\textwidth]{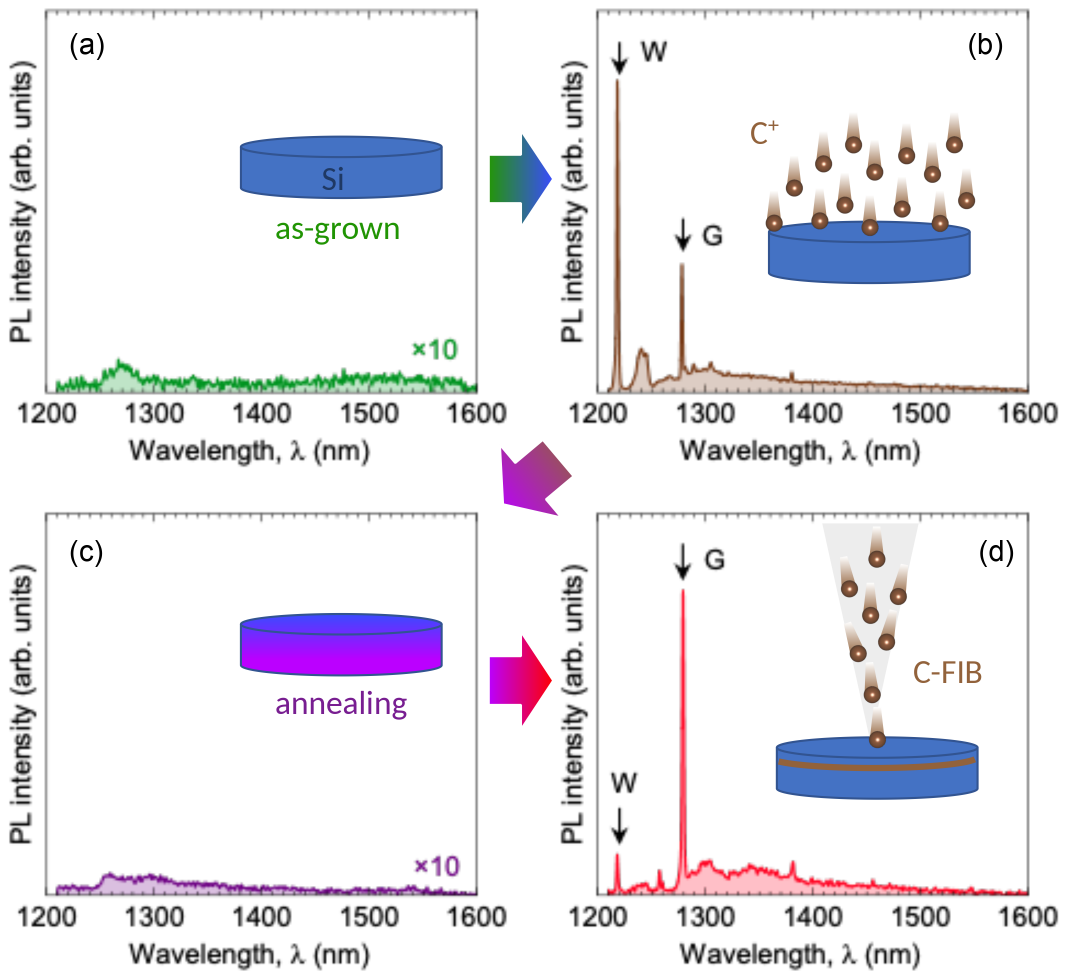}
\caption{Activation of G centers using multi-step implantation and annealing. 
(a) PL spectrum in an as-grown high-purity silicon wafer. 
(b) Step~1: PL spectrum after broad-beam implantation with $\mathrm{C^{+}}$ ions to a fluence of $1 \times 10^{12} \, \mathrm{cm^{-2}}$ at an acceleration potential of $13 \, \mathrm{kV}$. 
(c) Step~2: PL spectrum after annealing at a temperature of $500^{\circ}\mathrm{C}$ for two hours. 
(d) Step~3: PL spectrum from a locally activated area using C-FIB.} \label{fig4}
\end{figure}

To further increase the probability of the creation of G centers, we now apply a multi-step implantation and annealing protocol (Fig.~\ref{fig4}). 
A PL spectrum in the as-grown silicon wafer has some background due residual defects with energy states in the silicon bandgap is shown in Fig.~\ref{fig4}(a). 
We then perform a broad beam $\mathrm{C^{+}}$ implantation with an energy of $13 \, \mathrm{keV}$ (corresponding to the mean implantation depth of $46 \, \mathrm{nm}$) and a fluence $\Phi = 1 \times 10^{12} \, \mathrm{cm^{-2}}$. 
A PL spectrum after this implantation is presented in Fig.~\ref{fig4}(b). 
The ratio between the ZPL intensity of the G and W centers is similar to that after the C-FIB implantation of Fig.~\ref{fig2}(b) with higher energy, indicating that for the given range $I_{\mathrm{G}} / I_{\mathrm{W}} $ is independent of the implantation energy. 

To form the layer of the substitutional carbon pairs $\mathrm{C_{Si} - C_{Si}}$ and recover the silicon crystal structure after implantation damage, we perform annealing at a temperature of $500^{\circ}\mathrm{C}$ over two hours. 
Indeed, the PL spectrum shown in Fig.~\ref{fig4}(c) indicates the absence of the W and G centers in the PL spectrum. 
In the final step, we perform local activation of quantum emitters using our C-FIB. 
Due to possible thermally-induced diffusion of the implanted $\mathrm{C^{+}}$ ions (the first implantation step) during annealing, we use higher implantation energy of $20 \, \mathrm{keV}$. 
The corresponding mean implantation depth of $67 \, \mathrm{nm}$ and the in-depth distribution of $28 \, \mathrm{nm}$, but with a maximum of collision events around $45 \, \mathrm{nm}$ ensures the overlap of the first and second implantation layers.
A PL spectrum from an implanted spot is presented in Fig.~\ref{fig4}(d).
It shows the preferential formation of the G centers compared to the W centers $I_{\mathrm{G}} / I_{\mathrm{W}} > 7$ (Table~\ref{Probability_GW}). 

It is worth noting that there are two types of carbon-related emitters in silicon with the same spectral position of the ZPL but not identical optical properties, which are recently referred to as the G and G* centers \cite{arXiv:2402.07705}. 
Due to the observation of the E line at around $\lambda_{\mathrm{E}} = 1380 \, \mathrm{nm}$ in the PL spectrum of Fig.~\ref{fig4}(d), we associate the observed emission with the G centers but not with the G* centers. 

\begin{figure}[tbp]
\includegraphics[width=.47\textwidth]{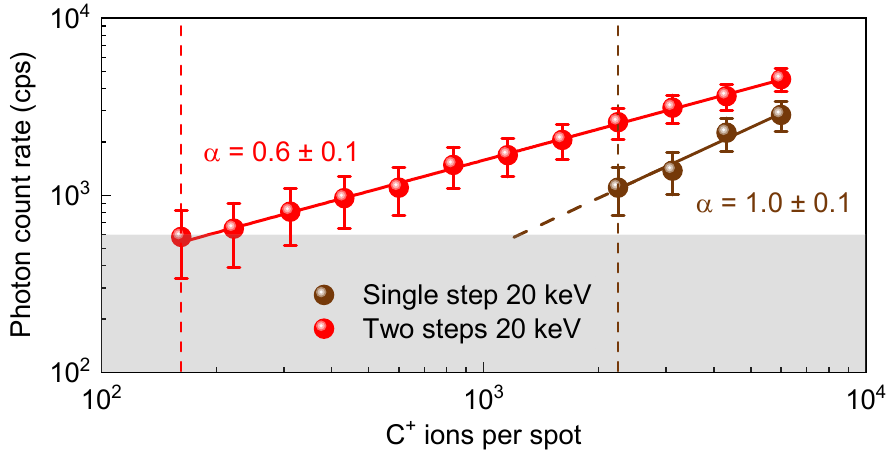}
\caption{Photon count rate from G centers as a function of the number of implanted $\mathrm{C^{+}}$ ions for single and two-step implantation protocols. 
The solid lines are fits to a power law $k n_{\mathrm{C}}^\alpha$. 
The dashed line is the extrapolation to the noise floor. 
The vertical thin dashed lines indicate the minimum number of implanted $\mathrm{C^{+}}$ ions required to detect the PL from G centers for different implantation protocols.} \label{fig5}
\end{figure}

A further advantage of the multi-step implantation is demonstrated in Fig.~\ref{fig5}, where the creation yield of the G center is compared for different protocols. 
We analyze the photon count rate from the G centers as a function of the number of implanted $\mathrm{C^{+}}$ ions. 
In case of the single-step implantation protocol with an energy of 20\,keV, the photon count rate, i.e., the number of created G centers, increases linearly with $n_{\mathrm{C}}$. 
We obtain the threshold value of about $[n_{\mathrm{C}}]_{min} \approx 2000$, corresponding to the creation yield of $1 / [n_{\mathrm{C}}]_{min} \approx 0.05 \%$. For the two-step implantation protocol, the photon count rate increases sub-linearly following $n_{\mathrm{C}}^{0.6} $, which is very similar to that observed in a carbon-rich silicon wafer \cite{10.1364/oe.397377}. 
The creation threshold of the G center decreases to $[n_{\mathrm{C}}]_{min} \approx 150$, corresponding to the increase of the creation yield by one order of magnitude to $1 / [n_{\mathrm{C}}]_{min} \approx 0.7 \%$. 
However, this value is still one order of magnitude lower than in the carbon-rich silicon after annealing \cite{10.1038/s41467-022-35051-5}. 
Further optimization of the multi-step implantation protocol, such as implantation fluence and temperature as well as annealing temperature and duration, is required, which is beyond the scope of this paper. 

In conclusion, the optimized CeC source formation protocol used in this work resulted in a useful $\mathrm{C^+}$ ion current of a few pA and a source lifetime of close to 100 hours. 
Contamination in particular by Al could be avoided using tantalum crucibles.
Due to the low mass of carbon, a high-purity ion beam without contamination can be achieved using a Wien filter.
These values are more than sufficient for the demonstrated application based on the implantation of only small numbers of ions in spatially confined areas. Using the C-FIB, we can locally activate two types of single-photon emitters in high-purity silicon with sub-100-nm
resolution. One type of emitter, the W center consisting of three interstitial Si atoms, is generated in the as-grown wafer. 
The combination of multi-step implantation and single-step annealing is used to programmatically activate the G center, a quantum emitter composed of one Si and two C atoms. 
For the multi-step fabrication protocol, the creation yield of the G centers per carbon ion is increased by about one order of magnitude compared to the single-step protocol. 

Apart from the G center in silicon, the carbon impurity has been identified as the source of visible single-photon emission from single $\mathrm{V_B V^-_N}$ defects in hBN \cite{10.1038/s41563-020-00850-y}. 
This is the only defect in 2D materials, on which the optically detected magnetic resonance (ODMR) on a single spin has been reported so far \cite{10.1038/s41467-022-28169-z}. 
Furthermore, the implantation of $\mathrm{C^{+}}$ ions is crucial for the creation of room-temperature single-spin qubits in SiC based on the PL6 defects with a high ODMR readout contrast \cite{10.1093/nsr/nwab122}. 
The creation, isolation and application of single TR12 centers in diamond for room-temperature vector magnetometry, i.e., alternative to the NV defect, has been also realized using $\mathrm{^{12}C}$ ion implantation \cite{10.1038/s41534-022-00566-8}. 
Therefore, our approach with C-FIB
is applicable for creating single-photon emitters and single-spin
sensors in various material platforms.

\section*{Experimental section}

\subsection*{Focused ion beam}
The focused ion beam implantations were carried out utilizing a commercial mass-separated Raith VELION FIB-SEM system. 
The system accelerates the ion with up to +35\,kV and is equipped with a Wien filter to select the ionic species according to their mass-to-charge ratio.
The implantation accuracy is in the range of the beam resolution of sub 10\,nm, depending on the energy, the mass of the ions and the ion current.
The samples can be positioned with high accuracy using a laser interferometer stage while avoiding ion beam exposure and unwanted implantations. 
The ion current is measured in a Faraday cup and varies less than 0.5\% per hour.
The ion beam is scanned over the sample with electrostatic multipole elements and each site of the sample is irradiated for a calculated short amount of time to control the local implantation fluence. 

\subsection*{Confocal microscope}

All measurements  reported in this work were performed using a home-built, low-temperature confocal microscope. 
The confocal imaging system is built-up based on a customized closed-cycle helium cryostat (attocube, attoDRY800). 
The incident laser 637~nm laser beam is adjusted by a variable optical alternator. Both, the laser excitation path and PL detection are fiber-coupled. 
A dichroic mirror is used to direct the laser onto the silicon sample, which is focused via an objective. 
The sample is located on top of a nanopositioning system and visualized by a fiber-coupled white light LED to locate the region of interest. 
The PL emerging from  quantum emitters and the portion of the laser beam reflected from the sample are separated from each other by a second dichroic mirror and longpass filter. 
The PL is detected either by superconducting nanowire single-photon detectors (SNSPD) or using an InGaAs photodiode array (PDA) for wavelength selective measurements. 
In the Hanbury Brown and Twiss (HBT) configuration, the PL signal is split by a fiber-coupled beam splitter in a 50/50 ratio in order to record the second-order autocorrelation function.

\begin{acknowledgments}
This article is based upon work from COST Action FIT4NANO CA19140, supported by COST (European Cooperation in Science and Technology). \url{https://www.cost.eu/} \url{https://www.fit4nano.eu}
Parts of this research were carried out at the Ion Beam Center (IBC) at the Helmholtz-Zentrum Dresden - Rossendorf e.~V., a member of the Helmholtz Association. 
\end{acknowledgments}

\bibliography{C-FIB_literature} 

\begin{thebibliography}{40}%
\makeatletter
\providecommand \@ifxundefined [1]{%
 \@ifx{#1\undefined}
}%
\providecommand \@ifnum [1]{%
 \ifnum #1\expandafter \@firstoftwo
 \else \expandafter \@secondoftwo
 \fi
}%
\providecommand \@ifx [1]{%
 \ifx #1\expandafter \@firstoftwo
 \else \expandafter \@secondoftwo
 \fi
}%
\providecommand \natexlab [1]{#1}%
\providecommand \enquote  [1]{``#1''}%
\providecommand \bibnamefont  [1]{#1}%
\providecommand \bibfnamefont [1]{#1}%
\providecommand \citenamefont [1]{#1}%
\providecommand \href@noop [0]{\@secondoftwo}%
\providecommand \href [0]{\begingroup \@sanitize@url \@href}%
\providecommand \@href[1]{\@@startlink{#1}\@@href}%
\providecommand \@@href[1]{\endgroup#1\@@endlink}%
\providecommand \@sanitize@url [0]{\catcode `\\12\catcode `\$12\catcode
  `\&12\catcode `\#12\catcode `\^12\catcode `\_12\catcode `\%12\relax}%
\providecommand \@@startlink[1]{}%
\providecommand \@@endlink[0]{}%
\providecommand \url  [0]{\begingroup\@sanitize@url \@url }%
\providecommand \@url [1]{\endgroup\@href {#1}{\urlprefix }}%
\providecommand \urlprefix  [0]{URL }%
\providecommand \Eprint [0]{\href }%
\providecommand \doibase [0]{https://doi.org/}%
\providecommand \selectlanguage [0]{\@gobble}%
\providecommand \bibinfo  [0]{\@secondoftwo}%
\providecommand \bibfield  [0]{\@secondoftwo}%
\providecommand \translation [1]{[#1]}%
\providecommand \BibitemOpen [0]{}%
\providecommand \bibitemStop [0]{}%
\providecommand \bibitemNoStop [0]{.\EOS\space}%
\providecommand \EOS [0]{\spacefactor3000\relax}%
\providecommand \BibitemShut  [1]{\csname bibitem#1\endcsname}%
\let\auto@bib@innerbib\@empty
\bibitem [{\citenamefont {Aharonovich}\ \emph {et~al.}(2016)\citenamefont
  {Aharonovich}, \citenamefont {Englund},\ and\ \citenamefont
  {Toth}}]{10.1038/nphoton.2016.186}%
  \BibitemOpen
  \bibfield  {author} {\bibinfo {author} {\bibfnamefont {I.}~\bibnamefont
  {Aharonovich}}, \bibinfo {author} {\bibfnamefont {D.}~\bibnamefont
  {Englund}},\ and\ \bibinfo {author} {\bibfnamefont {M.}~\bibnamefont
  {Toth}},\ }\bibfield  {title} {\bibinfo {title} {{Solid-state single-photon
  emitters}},\ }\href {https://doi.org/10.1038/nphoton.2016.186} {\bibfield
  {journal} {\bibinfo  {journal} {Nature Photonics}\ }\textbf {\bibinfo
  {volume} {10}},\ \bibinfo {pages} {631} (\bibinfo {year} {2016})}\BibitemShut
  {NoStop}%
\bibitem [{\citenamefont {Katsumi}\ \emph {et~al.}(2024)\citenamefont
  {Katsumi}, \citenamefont {Ota},\ and\ \citenamefont
  {Benyoucef}}]{10.1002/qute.202300423}%
  \BibitemOpen
  \bibfield  {author} {\bibinfo {author} {\bibfnamefont {R.}~\bibnamefont
  {Katsumi}}, \bibinfo {author} {\bibfnamefont {Y.}~\bibnamefont {Ota}},\ and\
  \bibinfo {author} {\bibfnamefont {M.}~\bibnamefont {Benyoucef}},\ }\bibfield
  {title} {\bibinfo {title} {{Telecom‐Band Quantum Dots Compatible with
  Silicon Photonics for Photonic Quantum Applications}},\ }\bibfield  {journal}
  {\bibinfo  {journal} {Advanced Quantum Technologies}\ }\href
  {https://doi.org/10.1002/qute.202300423} {10.1002/qute.202300423} (\bibinfo
  {year} {2024})\BibitemShut {NoStop}%
\bibitem [{\citenamefont {Ohshima}\ \emph {et~al.}(2018)\citenamefont
  {Ohshima}, \citenamefont {Satoh}, \citenamefont {Kraus}, \citenamefont
  {Astakhov}, \citenamefont {Dyakonov},\ and\ \citenamefont
  {Baranov}}]{10.1088/1361-6463/aad0ec}%
  \BibitemOpen
  \bibfield  {author} {\bibinfo {author} {\bibfnamefont {T.}~\bibnamefont
  {Ohshima}}, \bibinfo {author} {\bibfnamefont {T.}~\bibnamefont {Satoh}},
  \bibinfo {author} {\bibfnamefont {H.}~\bibnamefont {Kraus}}, \bibinfo
  {author} {\bibfnamefont {G.~V.}\ \bibnamefont {Astakhov}}, \bibinfo {author}
  {\bibfnamefont {V.}~\bibnamefont {Dyakonov}},\ and\ \bibinfo {author}
  {\bibfnamefont {P.~G.}\ \bibnamefont {Baranov}},\ }\bibfield  {title}
  {\bibinfo {title} {{Creation of silicon vacancy in silicon carbide by proton
  beam writing toward quantum sensing applications}},\ }\href
  {https://doi.org/10.1088/1361-6463/aad0ec} {\bibfield  {journal} {\bibinfo
  {journal} {Journal of Physics D: Applied Physics}\ }\textbf {\bibinfo
  {volume} {51}},\ \bibinfo {pages} {333002} (\bibinfo {year}
  {2018})}\BibitemShut {NoStop}%
\bibitem [{\citenamefont {Höflich}\ \emph {et~al.}(2023)\citenamefont
  {Höflich}, \citenamefont {Hobler}, \citenamefont {Allen}, \citenamefont
  {Wirtz}, \citenamefont {Rius}, \citenamefont {McElwee-White}, \citenamefont
  {Krasheninnikov}, \citenamefont {Schmidt}, \citenamefont {Utke},
  \citenamefont {Klingner}, \citenamefont {Osenberg}, \citenamefont {Córdoba},
  \citenamefont {Djurabekova}, \citenamefont {Manke}, \citenamefont {Moll},
  \citenamefont {Manoccio}, \citenamefont {De~Teresa}, \citenamefont
  {Bischoff}, \citenamefont {Michler}, \citenamefont {De~Castro}, \citenamefont
  {Delobbe}, \citenamefont {Dunne}, \citenamefont {Dobrovolskiy}, \citenamefont
  {Frese}, \citenamefont {Gölzhäuser}, \citenamefont {Mazarov}, \citenamefont
  {Koelle}, \citenamefont {Möller}, \citenamefont {Pérez-Murano},
  \citenamefont {Philipp}, \citenamefont {Vollnhals},\ and\ \citenamefont
  {Hlawacek}}]{10.1063/5.0162597}%
  \BibitemOpen
  \bibfield  {author} {\bibinfo {author} {\bibfnamefont {K.}~\bibnamefont
  {Höflich}}, \bibinfo {author} {\bibfnamefont {G.}~\bibnamefont {Hobler}},
  \bibinfo {author} {\bibfnamefont {F.~I.}\ \bibnamefont {Allen}}, \bibinfo
  {author} {\bibfnamefont {T.}~\bibnamefont {Wirtz}}, \bibinfo {author}
  {\bibfnamefont {G.}~\bibnamefont {Rius}}, \bibinfo {author} {\bibfnamefont
  {L.}~\bibnamefont {McElwee-White}}, \bibinfo {author} {\bibfnamefont {A.~V.}\
  \bibnamefont {Krasheninnikov}}, \bibinfo {author} {\bibfnamefont
  {M.}~\bibnamefont {Schmidt}}, \bibinfo {author} {\bibfnamefont
  {I.}~\bibnamefont {Utke}}, \bibinfo {author} {\bibfnamefont {N.}~\bibnamefont
  {Klingner}}, \bibinfo {author} {\bibfnamefont {M.}~\bibnamefont {Osenberg}},
  \bibinfo {author} {\bibfnamefont {R.}~\bibnamefont {Córdoba}}, \bibinfo
  {author} {\bibfnamefont {F.}~\bibnamefont {Djurabekova}}, \bibinfo {author}
  {\bibfnamefont {I.}~\bibnamefont {Manke}}, \bibinfo {author} {\bibfnamefont
  {P.}~\bibnamefont {Moll}}, \bibinfo {author} {\bibfnamefont {M.}~\bibnamefont
  {Manoccio}}, \bibinfo {author} {\bibfnamefont {J.~M.}\ \bibnamefont
  {De~Teresa}}, \bibinfo {author} {\bibfnamefont {L.}~\bibnamefont {Bischoff}},
  \bibinfo {author} {\bibfnamefont {J.}~\bibnamefont {Michler}}, \bibinfo
  {author} {\bibfnamefont {O.}~\bibnamefont {De~Castro}}, \bibinfo {author}
  {\bibfnamefont {A.}~\bibnamefont {Delobbe}}, \bibinfo {author} {\bibfnamefont
  {P.}~\bibnamefont {Dunne}}, \bibinfo {author} {\bibfnamefont {O.~V.}\
  \bibnamefont {Dobrovolskiy}}, \bibinfo {author} {\bibfnamefont
  {N.}~\bibnamefont {Frese}}, \bibinfo {author} {\bibfnamefont
  {A.}~\bibnamefont {Gölzhäuser}}, \bibinfo {author} {\bibfnamefont
  {P.}~\bibnamefont {Mazarov}}, \bibinfo {author} {\bibfnamefont
  {D.}~\bibnamefont {Koelle}}, \bibinfo {author} {\bibfnamefont
  {W.}~\bibnamefont {Möller}}, \bibinfo {author} {\bibfnamefont
  {F.}~\bibnamefont {Pérez-Murano}}, \bibinfo {author} {\bibfnamefont
  {P.}~\bibnamefont {Philipp}}, \bibinfo {author} {\bibfnamefont
  {F.}~\bibnamefont {Vollnhals}},\ and\ \bibinfo {author} {\bibfnamefont
  {G.}~\bibnamefont {Hlawacek}},\ }\bibfield  {title} {\bibinfo {title}
  {Roadmap for focused ion beam technologies},\ }\bibfield  {journal} {\bibinfo
   {journal} {Applied Physics Reviews}\ }\textbf {\bibinfo {volume} {10}},\
  \href {https://doi.org/10.1063/5.0162597} {10.1063/5.0162597} (\bibinfo
  {year} {2023})\BibitemShut {NoStop}%
\bibitem [{\citenamefont {Awschalom}\ \emph {et~al.}(2018)\citenamefont
  {Awschalom}, \citenamefont {Hanson}, \citenamefont {Wrachtrup},\ and\
  \citenamefont {Zhou}}]{10.1038/s41566-018-0232-2e8e}%
  \BibitemOpen
  \bibfield  {author} {\bibinfo {author} {\bibfnamefont {D.~D.}\ \bibnamefont
  {Awschalom}}, \bibinfo {author} {\bibfnamefont {R.}~\bibnamefont {Hanson}},
  \bibinfo {author} {\bibfnamefont {J.}~\bibnamefont {Wrachtrup}},\ and\
  \bibinfo {author} {\bibfnamefont {B.~B.}\ \bibnamefont {Zhou}},\ }\bibfield
  {title} {\bibinfo {title} {{Quantum technologies with optically interfaced
  solid-state spins}},\ }\href {https://doi.org/10.1038/s41566-018-0232-2}
  {\bibfield  {journal} {\bibinfo  {journal} {Nature Photonics}\ }\textbf
  {\bibinfo {volume} {12}},\ \bibinfo {pages} {516} (\bibinfo {year}
  {2018})}\BibitemShut {NoStop}%
\bibitem [{\citenamefont {O'Donnell}\ \emph {et~al.}(1983)\citenamefont
  {O'Donnell}, \citenamefont {Lee},\ and\ \citenamefont
  {Watkins}}]{10.1016/0378-4363(83)90256-5}%
  \BibitemOpen
  \bibfield  {author} {\bibinfo {author} {\bibfnamefont {K.}~\bibnamefont
  {O'Donnell}}, \bibinfo {author} {\bibfnamefont {K.}~\bibnamefont {Lee}},\
  and\ \bibinfo {author} {\bibfnamefont {G.}~\bibnamefont {Watkins}},\
  }\bibfield  {title} {\bibinfo {title} {{Origin of the 0.97 eV luminescence in
  irradiated silicon}},\ }\href {https://doi.org/10.1016/0378-4363(83)90256-5}
  {\bibfield  {journal} {\bibinfo  {journal} {Physica B+C}\ }\textbf {\bibinfo
  {volume} {116}},\ \bibinfo {pages} {258} (\bibinfo {year}
  {1983})}\BibitemShut {NoStop}%
\bibitem [{\citenamefont {Udvarhelyi}\ \emph {et~al.}(2021)\citenamefont
  {Udvarhelyi}, \citenamefont {Somogyi}, \citenamefont {Thiering},\ and\
  \citenamefont {Gali}}]{10.1103/physrevlett.127.196402}%
  \BibitemOpen
  \bibfield  {author} {\bibinfo {author} {\bibfnamefont {P.}~\bibnamefont
  {Udvarhelyi}}, \bibinfo {author} {\bibfnamefont {B.}~\bibnamefont {Somogyi}},
  \bibinfo {author} {\bibfnamefont {G.}~\bibnamefont {Thiering}},\ and\
  \bibinfo {author} {\bibfnamefont {A.}~\bibnamefont {Gali}},\ }\bibfield
  {title} {\bibinfo {title} {{Identification of a Telecom Wavelength Single
  Photon Emitter in Silicon}},\ }\href
  {https://doi.org/10.1103/physrevlett.127.196402} {\bibfield  {journal}
  {\bibinfo  {journal} {Physical Review Letters}\ }\textbf {\bibinfo {volume}
  {127}},\ \bibinfo {pages} {196402} (\bibinfo {year} {2021})}\BibitemShut
  {NoStop}%
\bibitem [{\citenamefont {Hollenbach}\ \emph {et~al.}(2020)\citenamefont
  {Hollenbach}, \citenamefont {Berencén}, \citenamefont {Kentsch},
  \citenamefont {Helm},\ and\ \citenamefont {Astakhov}}]{10.1364/oe.397377}%
  \BibitemOpen
  \bibfield  {author} {\bibinfo {author} {\bibfnamefont {M.}~\bibnamefont
  {Hollenbach}}, \bibinfo {author} {\bibfnamefont {Y.}~\bibnamefont
  {Berencén}}, \bibinfo {author} {\bibfnamefont {U.}~\bibnamefont {Kentsch}},
  \bibinfo {author} {\bibfnamefont {M.}~\bibnamefont {Helm}},\ and\ \bibinfo
  {author} {\bibfnamefont {G.~V.}\ \bibnamefont {Astakhov}},\ }\bibfield
  {title} {\bibinfo {title} {{Engineering telecom single-photon emitters in
  silicon for scalable quantum photonics}},\ }\href
  {https://doi.org/10.1364/oe.397377} {\bibfield  {journal} {\bibinfo
  {journal} {Optics Express}\ }\textbf {\bibinfo {volume} {28}},\ \bibinfo
  {pages} {26111} (\bibinfo {year} {2020})}\BibitemShut {NoStop}%
\bibitem [{\citenamefont {Redjem}\ \emph {et~al.}(2020)\citenamefont {Redjem},
  \citenamefont {Durand}, \citenamefont {Herzig}, \citenamefont {Benali},
  \citenamefont {Pezzagna}, \citenamefont {Meijer}, \citenamefont {Kuznetsov},
  \citenamefont {Nguyen}, \citenamefont {Cueff}, \citenamefont {Gérard},
  \citenamefont {Robert-Philip}, \citenamefont {Gil}, \citenamefont {Caliste},
  \citenamefont {Pochet}, \citenamefont {Abbarchi}, \citenamefont {Jacques},
  \citenamefont {Dréau},\ and\ \citenamefont
  {Cassabois}}]{10.1038/s41928-020-00499-0}%
  \BibitemOpen
  \bibfield  {author} {\bibinfo {author} {\bibfnamefont {W.}~\bibnamefont
  {Redjem}}, \bibinfo {author} {\bibfnamefont {A.}~\bibnamefont {Durand}},
  \bibinfo {author} {\bibfnamefont {T.}~\bibnamefont {Herzig}}, \bibinfo
  {author} {\bibfnamefont {A.}~\bibnamefont {Benali}}, \bibinfo {author}
  {\bibfnamefont {S.}~\bibnamefont {Pezzagna}}, \bibinfo {author}
  {\bibfnamefont {J.}~\bibnamefont {Meijer}}, \bibinfo {author} {\bibfnamefont
  {A.~Y.}\ \bibnamefont {Kuznetsov}}, \bibinfo {author} {\bibfnamefont {H.~S.}\
  \bibnamefont {Nguyen}}, \bibinfo {author} {\bibfnamefont {S.}~\bibnamefont
  {Cueff}}, \bibinfo {author} {\bibfnamefont {J.-M.}\ \bibnamefont {Gérard}},
  \bibinfo {author} {\bibfnamefont {I.}~\bibnamefont {Robert-Philip}}, \bibinfo
  {author} {\bibfnamefont {B.}~\bibnamefont {Gil}}, \bibinfo {author}
  {\bibfnamefont {D.}~\bibnamefont {Caliste}}, \bibinfo {author} {\bibfnamefont
  {P.}~\bibnamefont {Pochet}}, \bibinfo {author} {\bibfnamefont
  {M.}~\bibnamefont {Abbarchi}}, \bibinfo {author} {\bibfnamefont
  {V.}~\bibnamefont {Jacques}}, \bibinfo {author} {\bibfnamefont
  {A.}~\bibnamefont {Dréau}},\ and\ \bibinfo {author} {\bibfnamefont
  {G.}~\bibnamefont {Cassabois}},\ }\bibfield  {title} {\bibinfo {title}
  {{Single artificial atoms in silicon emitting at telecom wavelengths}},\
  }\href {https://doi.org/10.1038/s41928-020-00499-0} {\bibfield  {journal}
  {\bibinfo  {journal} {Nature Electronics}\ }\textbf {\bibinfo {volume} {3}},\
  \bibinfo {pages} {738} (\bibinfo {year} {2020})}\BibitemShut {NoStop}%
\bibitem [{\citenamefont {Hollenbach}\ \emph
  {et~al.}(2022{\natexlab{a}})\citenamefont {Hollenbach}, \citenamefont
  {Klingner}, \citenamefont {Jagtap}, \citenamefont {Bischoff}, \citenamefont
  {Fowley}, \citenamefont {Kentsch}, \citenamefont {Hlawacek}, \citenamefont
  {Erbe}, \citenamefont {Abrosimov}, \citenamefont {Helm}, \citenamefont
  {Berencén},\ and\ \citenamefont {Astakhov}}]{10.1038/s41467-022-35051-5}%
  \BibitemOpen
  \bibfield  {author} {\bibinfo {author} {\bibfnamefont {M.}~\bibnamefont
  {Hollenbach}}, \bibinfo {author} {\bibfnamefont {N.}~\bibnamefont
  {Klingner}}, \bibinfo {author} {\bibfnamefont {N.~S.}\ \bibnamefont
  {Jagtap}}, \bibinfo {author} {\bibfnamefont {L.}~\bibnamefont {Bischoff}},
  \bibinfo {author} {\bibfnamefont {C.}~\bibnamefont {Fowley}}, \bibinfo
  {author} {\bibfnamefont {U.}~\bibnamefont {Kentsch}}, \bibinfo {author}
  {\bibfnamefont {G.}~\bibnamefont {Hlawacek}}, \bibinfo {author}
  {\bibfnamefont {A.}~\bibnamefont {Erbe}}, \bibinfo {author} {\bibfnamefont
  {N.~V.}\ \bibnamefont {Abrosimov}}, \bibinfo {author} {\bibfnamefont
  {M.}~\bibnamefont {Helm}}, \bibinfo {author} {\bibfnamefont {Y.}~\bibnamefont
  {Berencén}},\ and\ \bibinfo {author} {\bibfnamefont {G.~V.}\ \bibnamefont
  {Astakhov}},\ }\bibfield  {title} {\bibinfo {title} {{Wafer-scale
  nanofabrication of telecom single-photon emitters in silicon}},\ }\href
  {https://doi.org/10.1038/s41467-022-35051-5} {\bibfield  {journal} {\bibinfo
  {journal} {Nature Communications}\ }\textbf {\bibinfo {volume} {13}},\
  \bibinfo {pages} {7683} (\bibinfo {year} {2022}{\natexlab{a}})}\BibitemShut
  {NoStop}%
\bibitem [{\citenamefont {Redjem}\ \emph {et~al.}(2023)\citenamefont {Redjem},
  \citenamefont {Zhiyenbayev}, \citenamefont {Qarony}, \citenamefont {Ivanov},
  \citenamefont {Papapanos}, \citenamefont {Liu}, \citenamefont {Jhuria},
  \citenamefont {Balushi}, \citenamefont {Dhuey}, \citenamefont {Schwartzberg},
  \citenamefont {Tan}, \citenamefont {Schenkel},\ and\ \citenamefont
  {Kanté}}]{10.1038/s41467-023-38559-6}%
  \BibitemOpen
  \bibfield  {author} {\bibinfo {author} {\bibfnamefont {W.}~\bibnamefont
  {Redjem}}, \bibinfo {author} {\bibfnamefont {Y.}~\bibnamefont {Zhiyenbayev}},
  \bibinfo {author} {\bibfnamefont {W.}~\bibnamefont {Qarony}}, \bibinfo
  {author} {\bibfnamefont {V.}~\bibnamefont {Ivanov}}, \bibinfo {author}
  {\bibfnamefont {C.}~\bibnamefont {Papapanos}}, \bibinfo {author}
  {\bibfnamefont {W.}~\bibnamefont {Liu}}, \bibinfo {author} {\bibfnamefont
  {K.}~\bibnamefont {Jhuria}}, \bibinfo {author} {\bibfnamefont {Z.~Y.~A.}\
  \bibnamefont {Balushi}}, \bibinfo {author} {\bibfnamefont {S.}~\bibnamefont
  {Dhuey}}, \bibinfo {author} {\bibfnamefont {A.}~\bibnamefont {Schwartzberg}},
  \bibinfo {author} {\bibfnamefont {L.~Z.}\ \bibnamefont {Tan}}, \bibinfo
  {author} {\bibfnamefont {T.}~\bibnamefont {Schenkel}},\ and\ \bibinfo
  {author} {\bibfnamefont {B.}~\bibnamefont {Kanté}},\ }\bibfield  {title}
  {\bibinfo {title} {{All-silicon quantum light source by embedding an atomic
  emissive center in a nanophotonic cavity}},\ }\href
  {https://doi.org/10.1038/s41467-023-38559-6} {\bibfield  {journal} {\bibinfo
  {journal} {Nature Communications}\ }\textbf {\bibinfo {volume} {14}},\
  \bibinfo {pages} {3321} (\bibinfo {year} {2023})}\BibitemShut {NoStop}%
\bibitem [{\citenamefont {Prabhu}\ \emph {et~al.}(2023)\citenamefont {Prabhu},
  \citenamefont {Errando-Herranz}, \citenamefont {Santis}, \citenamefont
  {Christen}, \citenamefont {Chen}, \citenamefont {Gerlach},\ and\
  \citenamefont {Englund}}]{10.1038/s41467-023-37655-x}%
  \BibitemOpen
  \bibfield  {author} {\bibinfo {author} {\bibfnamefont {M.}~\bibnamefont
  {Prabhu}}, \bibinfo {author} {\bibfnamefont {C.}~\bibnamefont
  {Errando-Herranz}}, \bibinfo {author} {\bibfnamefont {L.~D.}\ \bibnamefont
  {Santis}}, \bibinfo {author} {\bibfnamefont {I.}~\bibnamefont {Christen}},
  \bibinfo {author} {\bibfnamefont {C.}~\bibnamefont {Chen}}, \bibinfo {author}
  {\bibfnamefont {C.}~\bibnamefont {Gerlach}},\ and\ \bibinfo {author}
  {\bibfnamefont {D.}~\bibnamefont {Englund}},\ }\bibfield  {title} {\bibinfo
  {title} {{Individually addressable and spectrally programmable artificial
  atoms in silicon photonics}},\ }\href
  {https://doi.org/10.1038/s41467-023-37655-x} {\bibfield  {journal} {\bibinfo
  {journal} {Nature Communications}\ }\textbf {\bibinfo {volume} {14}},\
  \bibinfo {pages} {2380} (\bibinfo {year} {2023})}\BibitemShut {NoStop}%
\bibitem [{\citenamefont {Durand}\ \emph {et~al.}(2021)\citenamefont {Durand},
  \citenamefont {Baron}, \citenamefont {Redjem}, \citenamefont {Herzig},
  \citenamefont {Benali}, \citenamefont {Pezzagna}, \citenamefont {Meijer},
  \citenamefont {Kuznetsov}, \citenamefont {Gérard}, \citenamefont
  {Robert-Philip}, \citenamefont {Abbarchi}, \citenamefont {Jacques},
  \citenamefont {Cassabois},\ and\ \citenamefont
  {Dréau}}]{10.1103/physrevlett.126.083602}%
  \BibitemOpen
  \bibfield  {author} {\bibinfo {author} {\bibfnamefont {A.}~\bibnamefont
  {Durand}}, \bibinfo {author} {\bibfnamefont {Y.}~\bibnamefont {Baron}},
  \bibinfo {author} {\bibfnamefont {W.}~\bibnamefont {Redjem}}, \bibinfo
  {author} {\bibfnamefont {T.}~\bibnamefont {Herzig}}, \bibinfo {author}
  {\bibfnamefont {A.}~\bibnamefont {Benali}}, \bibinfo {author} {\bibfnamefont
  {S.}~\bibnamefont {Pezzagna}}, \bibinfo {author} {\bibfnamefont
  {J.}~\bibnamefont {Meijer}}, \bibinfo {author} {\bibfnamefont {A.~Y.}\
  \bibnamefont {Kuznetsov}}, \bibinfo {author} {\bibfnamefont {J.-M.}\
  \bibnamefont {Gérard}}, \bibinfo {author} {\bibfnamefont {I.}~\bibnamefont
  {Robert-Philip}}, \bibinfo {author} {\bibfnamefont {M.}~\bibnamefont
  {Abbarchi}}, \bibinfo {author} {\bibfnamefont {V.}~\bibnamefont {Jacques}},
  \bibinfo {author} {\bibfnamefont {G.}~\bibnamefont {Cassabois}},\ and\
  \bibinfo {author} {\bibfnamefont {A.}~\bibnamefont {Dréau}},\ }\bibfield
  {title} {\bibinfo {title} {{Broad Diversity of Near-Infrared Single-Photon
  Emitters in Silicon}},\ }\href
  {https://doi.org/10.1103/physrevlett.126.083602} {\bibfield  {journal}
  {\bibinfo  {journal} {Physical Review Letters}\ }\textbf {\bibinfo {volume}
  {126}},\ \bibinfo {pages} {083602} (\bibinfo {year} {2021})}\BibitemShut
  {NoStop}%
\bibitem [{\citenamefont {Hollenbach}\ \emph
  {et~al.}(2022{\natexlab{b}})\citenamefont {Hollenbach}, \citenamefont
  {Jagtap}, \citenamefont {Fowley}, \citenamefont {Baratech}, \citenamefont
  {Guardia-Arce}, \citenamefont {Kentsch}, \citenamefont {Eichler-Volf},
  \citenamefont {Abrosimov}, \citenamefont {Erbe}, \citenamefont {Shin},
  \citenamefont {Kim}, \citenamefont {Helm}, \citenamefont {Lee}, \citenamefont
  {Astakhov},\ and\ \citenamefont {Berencén}}]{10.1063/5.0094715}%
  \BibitemOpen
  \bibfield  {author} {\bibinfo {author} {\bibfnamefont {M.}~\bibnamefont
  {Hollenbach}}, \bibinfo {author} {\bibfnamefont {N.~S.}\ \bibnamefont
  {Jagtap}}, \bibinfo {author} {\bibfnamefont {C.}~\bibnamefont {Fowley}},
  \bibinfo {author} {\bibfnamefont {J.}~\bibnamefont {Baratech}}, \bibinfo
  {author} {\bibfnamefont {V.}~\bibnamefont {Guardia-Arce}}, \bibinfo {author}
  {\bibfnamefont {U.}~\bibnamefont {Kentsch}}, \bibinfo {author} {\bibfnamefont
  {A.}~\bibnamefont {Eichler-Volf}}, \bibinfo {author} {\bibfnamefont {N.~V.}\
  \bibnamefont {Abrosimov}}, \bibinfo {author} {\bibfnamefont {A.}~\bibnamefont
  {Erbe}}, \bibinfo {author} {\bibfnamefont {C.}~\bibnamefont {Shin}}, \bibinfo
  {author} {\bibfnamefont {H.}~\bibnamefont {Kim}}, \bibinfo {author}
  {\bibfnamefont {M.}~\bibnamefont {Helm}}, \bibinfo {author} {\bibfnamefont
  {W.}~\bibnamefont {Lee}}, \bibinfo {author} {\bibfnamefont {G.~V.}\
  \bibnamefont {Astakhov}},\ and\ \bibinfo {author} {\bibfnamefont
  {Y.}~\bibnamefont {Berencén}},\ }\bibfield  {title} {\bibinfo {title}
  {{Metal-assisted chemically etched silicon nanopillars hosting telecom photon
  emitters}},\ }\href {https://doi.org/10.1063/5.0094715} {\bibfield  {journal}
  {\bibinfo  {journal} {Journal of Applied Physics}\ }\textbf {\bibinfo
  {volume} {132}},\ \bibinfo {pages} {033101} (\bibinfo {year}
  {2022}{\natexlab{b}})}\BibitemShut {NoStop}%
\bibitem [{\citenamefont {Komza}\ \emph {et~al.}(2022)\citenamefont {Komza},
  \citenamefont {Samutpraphoot}, \citenamefont {Odeh}, \citenamefont {Tang},
  \citenamefont {Mathew}, \citenamefont {Chang}, \citenamefont {Song},
  \citenamefont {Kim}, \citenamefont {Xiong}, \citenamefont {Hautier},\ and\
  \citenamefont {Sipahigil}}]{arXiv:2211.09305}%
  \BibitemOpen
  \bibfield  {author} {\bibinfo {author} {\bibfnamefont {L.}~\bibnamefont
  {Komza}}, \bibinfo {author} {\bibfnamefont {P.}~\bibnamefont
  {Samutpraphoot}}, \bibinfo {author} {\bibfnamefont {M.}~\bibnamefont {Odeh}},
  \bibinfo {author} {\bibfnamefont {Y.-L.}\ \bibnamefont {Tang}}, \bibinfo
  {author} {\bibfnamefont {M.}~\bibnamefont {Mathew}}, \bibinfo {author}
  {\bibfnamefont {J.}~\bibnamefont {Chang}}, \bibinfo {author} {\bibfnamefont
  {H.}~\bibnamefont {Song}}, \bibinfo {author} {\bibfnamefont {M.-K.}\
  \bibnamefont {Kim}}, \bibinfo {author} {\bibfnamefont {Y.}~\bibnamefont
  {Xiong}}, \bibinfo {author} {\bibfnamefont {G.}~\bibnamefont {Hautier}},\
  and\ \bibinfo {author} {\bibfnamefont {A.}~\bibnamefont {Sipahigil}},\
  }\bibfield  {title} {\bibinfo {title} {{Indistinguishable photons from an
  artificial atom in silicon photonics}},\ }\href@noop {} {\bibfield  {journal}
  {\bibinfo  {journal} {arXiv:2211.09305}\ } (\bibinfo {year}
  {2022})}\BibitemShut {NoStop}%
\bibitem [{\citenamefont {Ristori}\ \emph {et~al.}(2024)\citenamefont
  {Ristori}, \citenamefont {Khoury}, \citenamefont {Salvalaglio}, \citenamefont
  {Filippatos}, \citenamefont {Amato}, \citenamefont {Herzig}, \citenamefont
  {Meijer}, \citenamefont {Pezzagna}, \citenamefont {Hannani}, \citenamefont
  {Bollani}, \citenamefont {Barri}, \citenamefont {Ruiz}, \citenamefont
  {Granchi}, \citenamefont {Intonti}, \citenamefont {Abbarchi},\ and\
  \citenamefont {Biccari}}]{10.1002/adom.202301608}%
  \BibitemOpen
  \bibfield  {author} {\bibinfo {author} {\bibfnamefont {A.}~\bibnamefont
  {Ristori}}, \bibinfo {author} {\bibfnamefont {M.}~\bibnamefont {Khoury}},
  \bibinfo {author} {\bibfnamefont {M.}~\bibnamefont {Salvalaglio}}, \bibinfo
  {author} {\bibfnamefont {A.}~\bibnamefont {Filippatos}}, \bibinfo {author}
  {\bibfnamefont {M.}~\bibnamefont {Amato}}, \bibinfo {author} {\bibfnamefont
  {T.}~\bibnamefont {Herzig}}, \bibinfo {author} {\bibfnamefont
  {J.}~\bibnamefont {Meijer}}, \bibinfo {author} {\bibfnamefont
  {S.}~\bibnamefont {Pezzagna}}, \bibinfo {author} {\bibfnamefont
  {D.}~\bibnamefont {Hannani}}, \bibinfo {author} {\bibfnamefont
  {M.}~\bibnamefont {Bollani}}, \bibinfo {author} {\bibfnamefont
  {C.}~\bibnamefont {Barri}}, \bibinfo {author} {\bibfnamefont {C.~M.}\
  \bibnamefont {Ruiz}}, \bibinfo {author} {\bibfnamefont {N.}~\bibnamefont
  {Granchi}}, \bibinfo {author} {\bibfnamefont {F.}~\bibnamefont {Intonti}},
  \bibinfo {author} {\bibfnamefont {M.}~\bibnamefont {Abbarchi}},\ and\
  \bibinfo {author} {\bibfnamefont {F.}~\bibnamefont {Biccari}},\ }\bibfield
  {title} {\bibinfo {title} {{Strain Engineering of the Electronic States of
  Silicon‐Based Quantum Emitters}},\ }\bibfield  {journal} {\bibinfo
  {journal} {Advanced Optical Materials}\ }\textbf {\bibinfo {volume} {12}},\
  \href {https://doi.org/10.1002/adom.202301608} {10.1002/adom.202301608}
  (\bibinfo {year} {2024})\BibitemShut {NoStop}%
\bibitem [{\citenamefont {Day}\ \emph {et~al.}(2023)\citenamefont {Day},
  \citenamefont {Sutula}, \citenamefont {Dietz}, \citenamefont {Raun},
  \citenamefont {Sukachev}, \citenamefont {Bhaskar},\ and\ \citenamefont
  {Hu}}]{arXiv:2311.08276}%
  \BibitemOpen
  \bibfield  {author} {\bibinfo {author} {\bibfnamefont {A.~M.}\ \bibnamefont
  {Day}}, \bibinfo {author} {\bibfnamefont {M.}~\bibnamefont {Sutula}},
  \bibinfo {author} {\bibfnamefont {J.~R.}\ \bibnamefont {Dietz}}, \bibinfo
  {author} {\bibfnamefont {A.}~\bibnamefont {Raun}}, \bibinfo {author}
  {\bibfnamefont {D.~D.}\ \bibnamefont {Sukachev}}, \bibinfo {author}
  {\bibfnamefont {M.~K.}\ \bibnamefont {Bhaskar}},\ and\ \bibinfo {author}
  {\bibfnamefont {E.~L.}\ \bibnamefont {Hu}},\ }\bibfield  {title} {\bibinfo
  {title} {{Electrical Manipulation of Telecom Color Centers in Silicon}},\
  }\href@noop {} {\bibfield  {journal} {\bibinfo  {journal} {arXiv:2311.08276}\
  } (\bibinfo {year} {2023})}\BibitemShut {NoStop}%
\bibitem [{\citenamefont {Lefaucher}\ \emph {et~al.}(2023)\citenamefont
  {Lefaucher}, \citenamefont {Jager}, \citenamefont {Calvo}, \citenamefont
  {Durand}, \citenamefont {Baron}, \citenamefont {Cache}, \citenamefont
  {Jacques}, \citenamefont {Robert-Philip}, \citenamefont {Cassabois},
  \citenamefont {Herzig}, \citenamefont {Meijer}, \citenamefont {Pezzagna},
  \citenamefont {Khoury}, \citenamefont {Abbarchi}, \citenamefont {Dréau},\
  and\ \citenamefont {Gérard}}]{10.1063/5.0130196}%
  \BibitemOpen
  \bibfield  {author} {\bibinfo {author} {\bibfnamefont {B.}~\bibnamefont
  {Lefaucher}}, \bibinfo {author} {\bibfnamefont {J.-B.}\ \bibnamefont
  {Jager}}, \bibinfo {author} {\bibfnamefont {V.}~\bibnamefont {Calvo}},
  \bibinfo {author} {\bibfnamefont {A.}~\bibnamefont {Durand}}, \bibinfo
  {author} {\bibfnamefont {Y.}~\bibnamefont {Baron}}, \bibinfo {author}
  {\bibfnamefont {F.}~\bibnamefont {Cache}}, \bibinfo {author} {\bibfnamefont
  {V.}~\bibnamefont {Jacques}}, \bibinfo {author} {\bibfnamefont
  {I.}~\bibnamefont {Robert-Philip}}, \bibinfo {author} {\bibfnamefont
  {G.}~\bibnamefont {Cassabois}}, \bibinfo {author} {\bibfnamefont
  {T.}~\bibnamefont {Herzig}}, \bibinfo {author} {\bibfnamefont
  {J.}~\bibnamefont {Meijer}}, \bibinfo {author} {\bibfnamefont
  {S.}~\bibnamefont {Pezzagna}}, \bibinfo {author} {\bibfnamefont
  {M.}~\bibnamefont {Khoury}}, \bibinfo {author} {\bibfnamefont
  {M.}~\bibnamefont {Abbarchi}}, \bibinfo {author} {\bibfnamefont
  {A.}~\bibnamefont {Dréau}},\ and\ \bibinfo {author} {\bibfnamefont {J.-M.}\
  \bibnamefont {Gérard}},\ }\bibfield  {title} {\bibinfo {title}
  {{Cavity-enhanced zero-phonon emission from an ensemble of G centers in a
  silicon-on-insulator microring}},\ }\href {https://doi.org/10.1063/5.0130196}
  {\bibfield  {journal} {\bibinfo  {journal} {Applied Physics Letters}\
  }\textbf {\bibinfo {volume} {122}},\ \bibinfo {pages} {061109} (\bibinfo
  {year} {2023})}\BibitemShut {NoStop}%
\bibitem [{\citenamefont {Saggio}\ \emph {et~al.}(2023)\citenamefont {Saggio},
  \citenamefont {Errando-Herranz}, \citenamefont {Gyger}, \citenamefont
  {Panuski}, \citenamefont {Prabhu}, \citenamefont {Santis}, \citenamefont
  {Christen}, \citenamefont {Ornelas-Huerta}, \citenamefont {Raniwala},
  \citenamefont {Gerlach}, \citenamefont {Colangelo},\ and\ \citenamefont
  {Englund}}]{arXiv:2302.10230}%
  \BibitemOpen
  \bibfield  {author} {\bibinfo {author} {\bibfnamefont {V.}~\bibnamefont
  {Saggio}}, \bibinfo {author} {\bibfnamefont {C.}~\bibnamefont
  {Errando-Herranz}}, \bibinfo {author} {\bibfnamefont {S.}~\bibnamefont
  {Gyger}}, \bibinfo {author} {\bibfnamefont {C.}~\bibnamefont {Panuski}},
  \bibinfo {author} {\bibfnamefont {M.}~\bibnamefont {Prabhu}}, \bibinfo
  {author} {\bibfnamefont {L.~D.}\ \bibnamefont {Santis}}, \bibinfo {author}
  {\bibfnamefont {I.}~\bibnamefont {Christen}}, \bibinfo {author}
  {\bibfnamefont {D.}~\bibnamefont {Ornelas-Huerta}}, \bibinfo {author}
  {\bibfnamefont {H.}~\bibnamefont {Raniwala}}, \bibinfo {author}
  {\bibfnamefont {C.}~\bibnamefont {Gerlach}}, \bibinfo {author} {\bibfnamefont
  {M.}~\bibnamefont {Colangelo}},\ and\ \bibinfo {author} {\bibfnamefont
  {D.}~\bibnamefont {Englund}},\ }\bibfield  {title} {\bibinfo {title}
  {{Cavity-enhanced single artificial atoms in silicon}},\ }\href@noop {}
  {\bibfield  {journal} {\bibinfo  {journal} {arXiv:2302.10230}\ } (\bibinfo
  {year} {2023})}\BibitemShut {NoStop}%
\bibitem [{\citenamefont {Berhanuddin}\ \emph {et~al.}(2012)\citenamefont
  {Berhanuddin}, \citenamefont {Lourenço}, \citenamefont {Gwilliam},\ and\
  \citenamefont {Homewood}}]{10.1002/adfm.201103034}%
  \BibitemOpen
  \bibfield  {author} {\bibinfo {author} {\bibfnamefont {D.~D.}\ \bibnamefont
  {Berhanuddin}}, \bibinfo {author} {\bibfnamefont {M.~A.}\ \bibnamefont
  {Lourenço}}, \bibinfo {author} {\bibfnamefont {R.~M.}\ \bibnamefont
  {Gwilliam}},\ and\ \bibinfo {author} {\bibfnamefont {K.~P.}\ \bibnamefont
  {Homewood}},\ }\bibfield  {title} {\bibinfo {title} {{Co‐Implantation of
  Carbon and Protons: An Integrated Silicon Device Technology Compatible Method
  to Generate the Lasing G‐Center}},\ }\href@noop {} {\bibfield  {journal}
  {\bibinfo  {journal} {Advanced Functional Materials}\ }\textbf {\bibinfo
  {volume} {22}},\ \bibinfo {pages} {2709} (\bibinfo {year}
  {2012})}\BibitemShut {NoStop}%
\bibitem [{\citenamefont {Beaufils}\ \emph {et~al.}(2018)\citenamefont
  {Beaufils}, \citenamefont {Redjem}, \citenamefont {Rousseau}, \citenamefont
  {Jacques}, \citenamefont {Kuznetsov}, \citenamefont {Raynaud}, \citenamefont
  {Voisin}, \citenamefont {Benali}, \citenamefont {Herzig}, \citenamefont
  {Pezzagna}, \citenamefont {Meijer}, \citenamefont {Abbarchi},\ and\
  \citenamefont {Cassabois}}]{10.1103/physrevb.97.035303}%
  \BibitemOpen
  \bibfield  {author} {\bibinfo {author} {\bibfnamefont {C.}~\bibnamefont
  {Beaufils}}, \bibinfo {author} {\bibfnamefont {W.}~\bibnamefont {Redjem}},
  \bibinfo {author} {\bibfnamefont {E.}~\bibnamefont {Rousseau}}, \bibinfo
  {author} {\bibfnamefont {V.}~\bibnamefont {Jacques}}, \bibinfo {author}
  {\bibfnamefont {A.~Y.}\ \bibnamefont {Kuznetsov}}, \bibinfo {author}
  {\bibfnamefont {C.}~\bibnamefont {Raynaud}}, \bibinfo {author} {\bibfnamefont
  {C.}~\bibnamefont {Voisin}}, \bibinfo {author} {\bibfnamefont
  {A.}~\bibnamefont {Benali}}, \bibinfo {author} {\bibfnamefont
  {T.}~\bibnamefont {Herzig}}, \bibinfo {author} {\bibfnamefont
  {S.}~\bibnamefont {Pezzagna}}, \bibinfo {author} {\bibfnamefont
  {J.}~\bibnamefont {Meijer}}, \bibinfo {author} {\bibfnamefont
  {M.}~\bibnamefont {Abbarchi}},\ and\ \bibinfo {author} {\bibfnamefont
  {G.}~\bibnamefont {Cassabois}},\ }\bibfield  {title} {\bibinfo {title}
  {{Optical properties of an ensemble of G-centers in silicon}},\ }\href@noop
  {} {\bibfield  {journal} {\bibinfo  {journal} {Physical Review B}\ }\textbf
  {\bibinfo {volume} {97}},\ \bibinfo {pages} {035303} (\bibinfo {year}
  {2018})}\BibitemShut {NoStop}%
\bibitem [{\citenamefont {Zhiyenbayev}\ \emph {et~al.}(2023)\citenamefont
  {Zhiyenbayev}, \citenamefont {Redjem}, \citenamefont {Ivanov}, \citenamefont
  {Qarony}, \citenamefont {Papapanos}, \citenamefont {Simoni}, \citenamefont
  {Liu}, \citenamefont {Jhuria}, \citenamefont {Tan}, \citenamefont
  {Schenkel},\ and\ \citenamefont {Kanté}}]{10.1364/oe.482311}%
  \BibitemOpen
  \bibfield  {author} {\bibinfo {author} {\bibfnamefont {Y.}~\bibnamefont
  {Zhiyenbayev}}, \bibinfo {author} {\bibfnamefont {W.}~\bibnamefont {Redjem}},
  \bibinfo {author} {\bibfnamefont {V.}~\bibnamefont {Ivanov}}, \bibinfo
  {author} {\bibfnamefont {W.}~\bibnamefont {Qarony}}, \bibinfo {author}
  {\bibfnamefont {C.}~\bibnamefont {Papapanos}}, \bibinfo {author}
  {\bibfnamefont {J.}~\bibnamefont {Simoni}}, \bibinfo {author} {\bibfnamefont
  {W.}~\bibnamefont {Liu}}, \bibinfo {author} {\bibfnamefont {K.}~\bibnamefont
  {Jhuria}}, \bibinfo {author} {\bibfnamefont {L.~Z.}\ \bibnamefont {Tan}},
  \bibinfo {author} {\bibfnamefont {T.}~\bibnamefont {Schenkel}},\ and\
  \bibinfo {author} {\bibfnamefont {B.}~\bibnamefont {Kanté}},\ }\bibfield
  {title} {\bibinfo {title} {{Scalable manufacturing of quantum light emitters
  in silicon under rapid thermal annealing}},\ }\href
  {https://doi.org/10.1364/oe.482311} {\bibfield  {journal} {\bibinfo
  {journal} {Optics Express}\ }\textbf {\bibinfo {volume} {31}},\ \bibinfo
  {pages} {8352} (\bibinfo {year} {2023})}\BibitemShut {NoStop}%
\bibitem [{\citenamefont {Baron}\ \emph
  {et~al.}(2022{\natexlab{a}})\citenamefont {Baron}, \citenamefont {Durand},
  \citenamefont {Herzig}, \citenamefont {Khoury}, \citenamefont {Pezzagna},
  \citenamefont {Meijer}, \citenamefont {Robert-Philip}, \citenamefont
  {Abbarchi}, \citenamefont {Hartmann}, \citenamefont {Reboh}, \citenamefont
  {Gérard}, \citenamefont {Jacques}, \citenamefont {Cassabois},\ and\
  \citenamefont {Dréau}}]{10.1063/5.0097407}%
  \BibitemOpen
  \bibfield  {author} {\bibinfo {author} {\bibfnamefont {Y.}~\bibnamefont
  {Baron}}, \bibinfo {author} {\bibfnamefont {A.}~\bibnamefont {Durand}},
  \bibinfo {author} {\bibfnamefont {T.}~\bibnamefont {Herzig}}, \bibinfo
  {author} {\bibfnamefont {M.}~\bibnamefont {Khoury}}, \bibinfo {author}
  {\bibfnamefont {S.}~\bibnamefont {Pezzagna}}, \bibinfo {author}
  {\bibfnamefont {J.}~\bibnamefont {Meijer}}, \bibinfo {author} {\bibfnamefont
  {I.}~\bibnamefont {Robert-Philip}}, \bibinfo {author} {\bibfnamefont
  {M.}~\bibnamefont {Abbarchi}}, \bibinfo {author} {\bibfnamefont {J.-M.}\
  \bibnamefont {Hartmann}}, \bibinfo {author} {\bibfnamefont {S.}~\bibnamefont
  {Reboh}}, \bibinfo {author} {\bibfnamefont {J.-M.}\ \bibnamefont {Gérard}},
  \bibinfo {author} {\bibfnamefont {V.}~\bibnamefont {Jacques}}, \bibinfo
  {author} {\bibfnamefont {G.}~\bibnamefont {Cassabois}},\ and\ \bibinfo
  {author} {\bibfnamefont {A.}~\bibnamefont {Dréau}},\ }\bibfield  {title}
  {\bibinfo {title} {{Single G centers in silicon fabricated by co-implantation
  with carbon and proton}},\ }\href {https://doi.org/10.1063/5.0097407}
  {\bibfield  {journal} {\bibinfo  {journal} {Applied Physics Letters}\
  }\textbf {\bibinfo {volume} {121}},\ \bibinfo {pages} {084003} (\bibinfo
  {year} {2022}{\natexlab{a}})}\BibitemShut {NoStop}%
\bibitem [{\citenamefont {Andrini}\ \emph {et~al.}(2024)\citenamefont
  {Andrini}, \citenamefont {Zanelli}, \citenamefont {Tchernij}, \citenamefont
  {Corte}, \citenamefont {Hernández}, \citenamefont {Verna}, \citenamefont
  {Cocuzza}, \citenamefont {Bernardi}, \citenamefont {Virzì}, \citenamefont
  {Traina}, \citenamefont {Degiovanni}, \citenamefont {Genovese}, \citenamefont
  {Olivero},\ and\ \citenamefont {Forneris}}]{10.1038/s43246-024-00486-4}%
  \BibitemOpen
  \bibfield  {author} {\bibinfo {author} {\bibfnamefont {G.}~\bibnamefont
  {Andrini}}, \bibinfo {author} {\bibfnamefont {G.}~\bibnamefont {Zanelli}},
  \bibinfo {author} {\bibfnamefont {S.~D.}\ \bibnamefont {Tchernij}}, \bibinfo
  {author} {\bibfnamefont {E.}~\bibnamefont {Corte}}, \bibinfo {author}
  {\bibfnamefont {E.~N.}\ \bibnamefont {Hernández}}, \bibinfo {author}
  {\bibfnamefont {A.}~\bibnamefont {Verna}}, \bibinfo {author} {\bibfnamefont
  {M.}~\bibnamefont {Cocuzza}}, \bibinfo {author} {\bibfnamefont
  {E.}~\bibnamefont {Bernardi}}, \bibinfo {author} {\bibfnamefont
  {S.}~\bibnamefont {Virzì}}, \bibinfo {author} {\bibfnamefont
  {P.}~\bibnamefont {Traina}}, \bibinfo {author} {\bibfnamefont {I.~P.}\
  \bibnamefont {Degiovanni}}, \bibinfo {author} {\bibfnamefont
  {M.}~\bibnamefont {Genovese}}, \bibinfo {author} {\bibfnamefont
  {P.}~\bibnamefont {Olivero}},\ and\ \bibinfo {author} {\bibfnamefont
  {J.}~\bibnamefont {Forneris}},\ }\bibfield  {title} {\bibinfo {title}
  {{Activation of telecom emitters in silicon upon ion implantation and ns
  pulsed laser annealing}},\ }\href
  {https://doi.org/10.1038/s43246-024-00486-4} {\bibfield  {journal} {\bibinfo
  {journal} {Communications Materials}\ }\textbf {\bibinfo {volume} {5}},\
  \bibinfo {pages} {47} (\bibinfo {year} {2024})}\BibitemShut {NoStop}%
\bibitem [{\citenamefont {Quard}\ \emph {et~al.}(2024)\citenamefont {Quard},
  \citenamefont {Khoury}, \citenamefont {Wang}, \citenamefont {Herzig},
  \citenamefont {Meijer}, \citenamefont {Pezzagna}, \citenamefont {Cueff},
  \citenamefont {Grojo}, \citenamefont {Abbarchi}, \citenamefont {Nguyen},
  \citenamefont {Chauvin},\ and\ \citenamefont
  {Wood}}]{10.1103/physrevapplied.21.044014}%
  \BibitemOpen
  \bibfield  {author} {\bibinfo {author} {\bibfnamefont {H.}~\bibnamefont
  {Quard}}, \bibinfo {author} {\bibfnamefont {M.}~\bibnamefont {Khoury}},
  \bibinfo {author} {\bibfnamefont {A.}~\bibnamefont {Wang}}, \bibinfo {author}
  {\bibfnamefont {T.}~\bibnamefont {Herzig}}, \bibinfo {author} {\bibfnamefont
  {J.}~\bibnamefont {Meijer}}, \bibinfo {author} {\bibfnamefont
  {S.}~\bibnamefont {Pezzagna}}, \bibinfo {author} {\bibfnamefont
  {S.}~\bibnamefont {Cueff}}, \bibinfo {author} {\bibfnamefont
  {D.}~\bibnamefont {Grojo}}, \bibinfo {author} {\bibfnamefont
  {M.}~\bibnamefont {Abbarchi}}, \bibinfo {author} {\bibfnamefont {H.~S.}\
  \bibnamefont {Nguyen}}, \bibinfo {author} {\bibfnamefont {N.}~\bibnamefont
  {Chauvin}},\ and\ \bibinfo {author} {\bibfnamefont {T.}~\bibnamefont
  {Wood}},\ }\bibfield  {title} {\bibinfo {title} {{Femtosecond-laser-induced
  creation of G and W color centers in silicon-on-insulator substrates}},\
  }\href {https://doi.org/10.1103/physrevapplied.21.044014} {\bibfield
  {journal} {\bibinfo  {journal} {Physical Review Applied}\ }\textbf {\bibinfo
  {volume} {21}},\ \bibinfo {pages} {044014} (\bibinfo {year}
  {2024})}\BibitemShut {NoStop}%
\bibitem [{\citenamefont {Jhuria}\ \emph {et~al.}(2023)\citenamefont {Jhuria},
  \citenamefont {Ivanov}, \citenamefont {Polley}, \citenamefont {Liu},
  \citenamefont {Persaud}, \citenamefont {Zhiyenbayev}, \citenamefont {Redjem},
  \citenamefont {Qarony}, \citenamefont {Parajuli}, \citenamefont {Ji},
  \citenamefont {Gonsalves}, \citenamefont {Bokor}, \citenamefont {Tan},
  \citenamefont {Kante},\ and\ \citenamefont {Schenkel}}]{arXiv:2307.05759}%
  \BibitemOpen
  \bibfield  {author} {\bibinfo {author} {\bibfnamefont {K.}~\bibnamefont
  {Jhuria}}, \bibinfo {author} {\bibfnamefont {V.}~\bibnamefont {Ivanov}},
  \bibinfo {author} {\bibfnamefont {D.}~\bibnamefont {Polley}}, \bibinfo
  {author} {\bibfnamefont {W.}~\bibnamefont {Liu}}, \bibinfo {author}
  {\bibfnamefont {A.}~\bibnamefont {Persaud}}, \bibinfo {author} {\bibfnamefont
  {Y.}~\bibnamefont {Zhiyenbayev}}, \bibinfo {author} {\bibfnamefont
  {W.}~\bibnamefont {Redjem}}, \bibinfo {author} {\bibfnamefont
  {W.}~\bibnamefont {Qarony}}, \bibinfo {author} {\bibfnamefont
  {P.}~\bibnamefont {Parajuli}}, \bibinfo {author} {\bibfnamefont
  {Q.}~\bibnamefont {Ji}}, \bibinfo {author} {\bibfnamefont {A.~J.}\
  \bibnamefont {Gonsalves}}, \bibinfo {author} {\bibfnamefont {J.}~\bibnamefont
  {Bokor}}, \bibinfo {author} {\bibfnamefont {L.~Z.}\ \bibnamefont {Tan}},
  \bibinfo {author} {\bibfnamefont {B.}~\bibnamefont {Kante}},\ and\ \bibinfo
  {author} {\bibfnamefont {T.}~\bibnamefont {Schenkel}},\ }\bibfield  {title}
  {\bibinfo {title} {{Programmable quantum emitter formation in silicon}},\
  }\href@noop {} {\bibfield  {journal} {\bibinfo  {journal} {arXiv:2307.05759}\
  } (\bibinfo {year} {2023})}\BibitemShut {NoStop}%
\bibitem [{\citenamefont {Chartrand}\ \emph {et~al.}(2018)\citenamefont
  {Chartrand}, \citenamefont {Bergeron}, \citenamefont {Morse}, \citenamefont
  {Riemann}, \citenamefont {Abrosimov}, \citenamefont {Becker}, \citenamefont
  {Pohl}, \citenamefont {Simmons},\ and\ \citenamefont
  {Thewalt}}]{10.1103/physrevb.98.195201}%
  \BibitemOpen
  \bibfield  {author} {\bibinfo {author} {\bibfnamefont {C.}~\bibnamefont
  {Chartrand}}, \bibinfo {author} {\bibfnamefont {L.}~\bibnamefont {Bergeron}},
  \bibinfo {author} {\bibfnamefont {K.~J.}\ \bibnamefont {Morse}}, \bibinfo
  {author} {\bibfnamefont {H.}~\bibnamefont {Riemann}}, \bibinfo {author}
  {\bibfnamefont {N.~V.}\ \bibnamefont {Abrosimov}}, \bibinfo {author}
  {\bibfnamefont {P.}~\bibnamefont {Becker}}, \bibinfo {author} {\bibfnamefont
  {H.-J.}\ \bibnamefont {Pohl}}, \bibinfo {author} {\bibfnamefont
  {S.}~\bibnamefont {Simmons}},\ and\ \bibinfo {author} {\bibfnamefont
  {M.~L.~W.}\ \bibnamefont {Thewalt}},\ }\bibfield  {title} {\bibinfo {title}
  {{Highly enriched Si-28 reveals remarkable optical linewidths and fine
  structure for well-known damage centers}},\ }\href@noop {} {\bibfield
  {journal} {\bibinfo  {journal} {Physical Review B}\ }\textbf {\bibinfo
  {volume} {98}},\ \bibinfo {pages} {195201} (\bibinfo {year}
  {2018})}\BibitemShut {NoStop}%
\bibitem [{\citenamefont {Mazarov}\ \emph {et~al.}(2009)\citenamefont
  {Mazarov}, \citenamefont {Wieck}, \citenamefont {Bischoff},\ and\
  \citenamefont {Pilz}}]{10.1116/1.3253471}%
  \BibitemOpen
  \bibfield  {author} {\bibinfo {author} {\bibfnamefont {P.}~\bibnamefont
  {Mazarov}}, \bibinfo {author} {\bibfnamefont {A.~D.}\ \bibnamefont {Wieck}},
  \bibinfo {author} {\bibfnamefont {L.}~\bibnamefont {Bischoff}},\ and\
  \bibinfo {author} {\bibfnamefont {W.}~\bibnamefont {Pilz}},\ }\bibfield
  {title} {\bibinfo {title} {Alloy liquid metal ion source for carbon focused
  ion beams},\ }\href {https://doi.org/10.1116/1.3253471} {\bibfield  {journal}
  {\bibinfo  {journal} {Journal of Vacuum Science {\&} Technology B:
  Microelectronics and Nanometer Structures}\ }\textbf {\bibinfo {volume}
  {27}},\ \bibinfo {pages} {L47} (\bibinfo {year} {2009})}\BibitemShut
  {NoStop}%
\bibitem [{\citenamefont {Okamoto}\ \emph {et~al.}(1990)\citenamefont
  {Okamoto}, \citenamefont {Massalski} \emph
  {et~al.}}]{10.31399/asm.hb.v03.a0006247}%
  \BibitemOpen
  \bibfield  {author} {\bibinfo {author} {\bibfnamefont {H.}~\bibnamefont
  {Okamoto}}, \bibinfo {author} {\bibfnamefont {T.}~\bibnamefont {Massalski}},
  \emph {et~al.},\ }\bibfield  {title} {\bibinfo {title} {Binary alloy phase
  diagrams},\ }\href@noop {} {\bibfield  {journal} {\bibinfo  {journal} {ASM
  International, Materials Park, OH, USA}\ }\textbf {\bibinfo {volume} {12}},\
  \bibinfo {pages} {3528} (\bibinfo {year} {1990})}\BibitemShut {NoStop}%
\bibitem [{\citenamefont {Alcock}\ \emph {et~al.}(1984)\citenamefont {Alcock},
  \citenamefont {Itkin},\ and\ \citenamefont
  {Horrigan}}]{10.1179/cmq.1984.23.3.309}%
  \BibitemOpen
  \bibfield  {author} {\bibinfo {author} {\bibfnamefont {C.~B.}\ \bibnamefont
  {Alcock}}, \bibinfo {author} {\bibfnamefont {V.~P.}\ \bibnamefont {Itkin}},\
  and\ \bibinfo {author} {\bibfnamefont {M.~K.}\ \bibnamefont {Horrigan}},\
  }\bibfield  {title} {\bibinfo {title} {Vapour pressure equations for the
  metallic elements: 298–2500k},\ }\href
  {https://doi.org/10.1179/cmq.1984.23.3.309} {\bibfield  {journal} {\bibinfo
  {journal} {Canadian Metallurgical Quarterly}\ }\textbf {\bibinfo {volume}
  {23}},\ \bibinfo {pages} {309} (\bibinfo {year} {1984})}\BibitemShut
  {NoStop}%
\bibitem [{\citenamefont {Buckley}\ \emph {et~al.}(2020)\citenamefont
  {Buckley}, \citenamefont {Tait}, \citenamefont {Moody}, \citenamefont
  {Primavera}, \citenamefont {Olson}, \citenamefont {Herman}, \citenamefont
  {Silverman}, \citenamefont {Rao}, \citenamefont {Nam}, \citenamefont
  {Mirin},\ and\ \citenamefont {Shainline}}]{10.1364/oe.386450}%
  \BibitemOpen
  \bibfield  {author} {\bibinfo {author} {\bibfnamefont {S.~M.}\ \bibnamefont
  {Buckley}}, \bibinfo {author} {\bibfnamefont {A.~N.}\ \bibnamefont {Tait}},
  \bibinfo {author} {\bibfnamefont {G.}~\bibnamefont {Moody}}, \bibinfo
  {author} {\bibfnamefont {B.}~\bibnamefont {Primavera}}, \bibinfo {author}
  {\bibfnamefont {S.}~\bibnamefont {Olson}}, \bibinfo {author} {\bibfnamefont
  {J.}~\bibnamefont {Herman}}, \bibinfo {author} {\bibfnamefont {K.~L.}\
  \bibnamefont {Silverman}}, \bibinfo {author} {\bibfnamefont {S.~P.}\
  \bibnamefont {Rao}}, \bibinfo {author} {\bibfnamefont {S.~W.}\ \bibnamefont
  {Nam}}, \bibinfo {author} {\bibfnamefont {R.~P.}\ \bibnamefont {Mirin}},\
  and\ \bibinfo {author} {\bibfnamefont {J.~M.}\ \bibnamefont {Shainline}},\
  }\bibfield  {title} {\bibinfo {title} {{Optimization of photoluminescence
  from W centers in silicon-on-insulator}},\ }\href@noop {} {\bibfield
  {journal} {\bibinfo  {journal} {Optics Express}\ }\textbf {\bibinfo {volume}
  {28}},\ \bibinfo {pages} {16057} (\bibinfo {year} {2020})}\BibitemShut
  {NoStop}%
\bibitem [{\citenamefont {Baron}\ \emph
  {et~al.}(2022{\natexlab{b}})\citenamefont {Baron}, \citenamefont {Durand},
  \citenamefont {Udvarhelyi}, \citenamefont {Herzig}, \citenamefont {Khoury},
  \citenamefont {Pezzagna}, \citenamefont {Meijer}, \citenamefont
  {Robert-Philip}, \citenamefont {Abbarchi}, \citenamefont {Hartmann},
  \citenamefont {Mazzocchi}, \citenamefont {G\'{e}rard}, \citenamefont {Gali},
  \citenamefont {Jacques}, \citenamefont {Cassabois},\ and\ \citenamefont
  {Dr\'{e}au}}]{10.1021/acsphotonics.2c00336}%
  \BibitemOpen
  \bibfield  {author} {\bibinfo {author} {\bibfnamefont {Y.}~\bibnamefont
  {Baron}}, \bibinfo {author} {\bibfnamefont {A.}~\bibnamefont {Durand}},
  \bibinfo {author} {\bibfnamefont {P.}~\bibnamefont {Udvarhelyi}}, \bibinfo
  {author} {\bibfnamefont {T.}~\bibnamefont {Herzig}}, \bibinfo {author}
  {\bibfnamefont {M.}~\bibnamefont {Khoury}}, \bibinfo {author} {\bibfnamefont
  {S.}~\bibnamefont {Pezzagna}}, \bibinfo {author} {\bibfnamefont
  {J.}~\bibnamefont {Meijer}}, \bibinfo {author} {\bibfnamefont
  {I.}~\bibnamefont {Robert-Philip}}, \bibinfo {author} {\bibfnamefont
  {M.}~\bibnamefont {Abbarchi}}, \bibinfo {author} {\bibfnamefont {J.-M.}\
  \bibnamefont {Hartmann}}, \bibinfo {author} {\bibfnamefont {V.}~\bibnamefont
  {Mazzocchi}}, \bibinfo {author} {\bibfnamefont {J.-M.}\ \bibnamefont
  {G\'{e}rard}}, \bibinfo {author} {\bibfnamefont {A.}~\bibnamefont {Gali}},
  \bibinfo {author} {\bibfnamefont {V.}~\bibnamefont {Jacques}}, \bibinfo
  {author} {\bibfnamefont {G.}~\bibnamefont {Cassabois}},\ and\ \bibinfo
  {author} {\bibfnamefont {A.}~\bibnamefont {Dr\'{e}au}},\ }\bibfield  {title}
  {\bibinfo {title} {{Detection of Single W-Centers in Silicon}},\ }\href@noop
  {} {\bibfield  {journal} {\bibinfo  {journal} {ACS Photonics}\ }\textbf
  {\bibinfo {volume} {9}},\ \bibinfo {pages} {2337} (\bibinfo {year}
  {2022}{\natexlab{b}})}\BibitemShut {NoStop}%
\bibitem [{\citenamefont {Deák}\ \emph {et~al.}(2023)\citenamefont {Deák},
  \citenamefont {Udvarhelyi}, \citenamefont {Thiering},\ and\ \citenamefont
  {Gali}}]{10.1038/s41467-023-36090-2}%
  \BibitemOpen
  \bibfield  {author} {\bibinfo {author} {\bibfnamefont {P.}~\bibnamefont
  {Deák}}, \bibinfo {author} {\bibfnamefont {P.}~\bibnamefont {Udvarhelyi}},
  \bibinfo {author} {\bibfnamefont {G.}~\bibnamefont {Thiering}},\ and\
  \bibinfo {author} {\bibfnamefont {A.}~\bibnamefont {Gali}},\ }\bibfield
  {title} {\bibinfo {title} {{The kinetics of carbon pair formation in silicon
  prohibits reaching thermal equilibrium}},\ }\href@noop {} {\bibfield
  {journal} {\bibinfo  {journal} {Nature Communications}\ }\textbf {\bibinfo
  {volume} {14}},\ \bibinfo {pages} {361} (\bibinfo {year} {2023})}\BibitemShut
  {NoStop}%
\bibitem [{\citenamefont {Liu}\ \emph {et~al.}(2023)\citenamefont {Liu},
  \citenamefont {Ivanov}, \citenamefont {Jhuria}, \citenamefont {Ji},
  \citenamefont {Persaud}, \citenamefont {Redjem}, \citenamefont {Simoni},
  \citenamefont {Zhiyenbayev}, \citenamefont {Kante}, \citenamefont {Lopez},
  \citenamefont {Tan},\ and\ \citenamefont {Schenkel}}]{arXiv:2302.05814}%
  \BibitemOpen
  \bibfield  {author} {\bibinfo {author} {\bibfnamefont {W.}~\bibnamefont
  {Liu}}, \bibinfo {author} {\bibfnamefont {V.}~\bibnamefont {Ivanov}},
  \bibinfo {author} {\bibfnamefont {K.}~\bibnamefont {Jhuria}}, \bibinfo
  {author} {\bibfnamefont {Q.}~\bibnamefont {Ji}}, \bibinfo {author}
  {\bibfnamefont {A.}~\bibnamefont {Persaud}}, \bibinfo {author} {\bibfnamefont
  {W.}~\bibnamefont {Redjem}}, \bibinfo {author} {\bibfnamefont
  {J.}~\bibnamefont {Simoni}}, \bibinfo {author} {\bibfnamefont
  {Y.}~\bibnamefont {Zhiyenbayev}}, \bibinfo {author} {\bibfnamefont
  {B.}~\bibnamefont {Kante}}, \bibinfo {author} {\bibfnamefont {J.~G.}\
  \bibnamefont {Lopez}}, \bibinfo {author} {\bibfnamefont {L.~Z.}\ \bibnamefont
  {Tan}},\ and\ \bibinfo {author} {\bibfnamefont {T.}~\bibnamefont
  {Schenkel}},\ }\bibfield  {title} {\bibinfo {title} {{Quantum emitter
  formation dynamics and probing of radiation induced atomic disorder in
  silicon}},\ }\href@noop {} {\bibfield  {journal} {\bibinfo  {journal}
  {arXiv:2302.05814}\ } (\bibinfo {year} {2023})}\BibitemShut {NoStop}%
\bibitem [{\citenamefont {Fuchs}\ \emph {et~al.}(2015)\citenamefont {Fuchs},
  \citenamefont {Stender}, \citenamefont {Trupke}, \citenamefont {Simin},
  \citenamefont {Pflaum}, \citenamefont {Dyakonov},\ and\ \citenamefont
  {Astakhov}}]{10.1038/ncomms8578}%
  \BibitemOpen
  \bibfield  {author} {\bibinfo {author} {\bibfnamefont {F.}~\bibnamefont
  {Fuchs}}, \bibinfo {author} {\bibfnamefont {B.}~\bibnamefont {Stender}},
  \bibinfo {author} {\bibfnamefont {M.}~\bibnamefont {Trupke}}, \bibinfo
  {author} {\bibfnamefont {D.}~\bibnamefont {Simin}}, \bibinfo {author}
  {\bibfnamefont {J.}~\bibnamefont {Pflaum}}, \bibinfo {author} {\bibfnamefont
  {V.}~\bibnamefont {Dyakonov}},\ and\ \bibinfo {author} {\bibfnamefont
  {G.~V.}\ \bibnamefont {Astakhov}},\ }\bibfield  {title} {\bibinfo {title}
  {{Engineering near-infrared single-photon emitters with optically active
  spins in ultrapure silicon carbide}},\ }\href@noop {} {\bibfield  {journal}
  {\bibinfo  {journal} {Nature Communications}\ }\textbf {\bibinfo {volume}
  {6}},\ \bibinfo {pages} {7578} (\bibinfo {year} {2015})}\BibitemShut
  {NoStop}%
\bibitem [{\citenamefont {Durand}\ \emph {et~al.}(2024)\citenamefont {Durand},
  \citenamefont {Baron}, \citenamefont {Cache}, \citenamefont {Herzig},
  \citenamefont {Khoury}, \citenamefont {Pezzagna}, \citenamefont {Meijer},
  \citenamefont {Hartmann}, \citenamefont {Reboh}, \citenamefont {Abbarchi},
  \citenamefont {Robert-Philip}, \citenamefont {Gérard}, \citenamefont
  {Jacques}, \citenamefont {Cassabois},\ and\ \citenamefont
  {Dréau}}]{arXiv:2402.07705}%
  \BibitemOpen
  \bibfield  {author} {\bibinfo {author} {\bibfnamefont {A.}~\bibnamefont
  {Durand}}, \bibinfo {author} {\bibfnamefont {Y.}~\bibnamefont {Baron}},
  \bibinfo {author} {\bibfnamefont {F.}~\bibnamefont {Cache}}, \bibinfo
  {author} {\bibfnamefont {T.}~\bibnamefont {Herzig}}, \bibinfo {author}
  {\bibfnamefont {M.}~\bibnamefont {Khoury}}, \bibinfo {author} {\bibfnamefont
  {S.}~\bibnamefont {Pezzagna}}, \bibinfo {author} {\bibfnamefont
  {J.}~\bibnamefont {Meijer}}, \bibinfo {author} {\bibfnamefont {J.-M.}\
  \bibnamefont {Hartmann}}, \bibinfo {author} {\bibfnamefont {S.}~\bibnamefont
  {Reboh}}, \bibinfo {author} {\bibfnamefont {M.}~\bibnamefont {Abbarchi}},
  \bibinfo {author} {\bibfnamefont {I.}~\bibnamefont {Robert-Philip}}, \bibinfo
  {author} {\bibfnamefont {J.-M.}\ \bibnamefont {Gérard}}, \bibinfo {author}
  {\bibfnamefont {V.}~\bibnamefont {Jacques}}, \bibinfo {author} {\bibfnamefont
  {G.}~\bibnamefont {Cassabois}},\ and\ \bibinfo {author} {\bibfnamefont
  {A.}~\bibnamefont {Dréau}},\ }\bibfield  {title} {\bibinfo {title} {{Genuine
  and faux single G centers in carbon-implanted silicon}},\ }\href@noop {}
  {\bibfield  {journal} {\bibinfo  {journal} {arXiv:2402.07705}\ } (\bibinfo
  {year} {2024})}\BibitemShut {NoStop}%
\bibitem [{\citenamefont {Mendelson}\ \emph {et~al.}(2021)\citenamefont
  {Mendelson}, \citenamefont {Chugh}, \citenamefont {Reimers}, \citenamefont
  {Cheng}, \citenamefont {Gottscholl}, \citenamefont {Long}, \citenamefont
  {Mellor}, \citenamefont {Zettl}, \citenamefont {Dyakonov}, \citenamefont
  {Beton}, \citenamefont {Novikov}, \citenamefont {Jagadish}, \citenamefont
  {Tan}, \citenamefont {Ford}, \citenamefont {Toth}, \citenamefont {Bradac},\
  and\ \citenamefont {Aharonovich}}]{10.1038/s41563-020-00850-y}%
  \BibitemOpen
  \bibfield  {author} {\bibinfo {author} {\bibfnamefont {N.}~\bibnamefont
  {Mendelson}}, \bibinfo {author} {\bibfnamefont {D.}~\bibnamefont {Chugh}},
  \bibinfo {author} {\bibfnamefont {J.~R.}\ \bibnamefont {Reimers}}, \bibinfo
  {author} {\bibfnamefont {T.~S.}\ \bibnamefont {Cheng}}, \bibinfo {author}
  {\bibfnamefont {A.}~\bibnamefont {Gottscholl}}, \bibinfo {author}
  {\bibfnamefont {H.}~\bibnamefont {Long}}, \bibinfo {author} {\bibfnamefont
  {C.~J.}\ \bibnamefont {Mellor}}, \bibinfo {author} {\bibfnamefont
  {A.}~\bibnamefont {Zettl}}, \bibinfo {author} {\bibfnamefont
  {V.}~\bibnamefont {Dyakonov}}, \bibinfo {author} {\bibfnamefont {P.~H.}\
  \bibnamefont {Beton}}, \bibinfo {author} {\bibfnamefont {S.~V.}\ \bibnamefont
  {Novikov}}, \bibinfo {author} {\bibfnamefont {C.}~\bibnamefont {Jagadish}},
  \bibinfo {author} {\bibfnamefont {H.~H.}\ \bibnamefont {Tan}}, \bibinfo
  {author} {\bibfnamefont {M.~J.}\ \bibnamefont {Ford}}, \bibinfo {author}
  {\bibfnamefont {M.}~\bibnamefont {Toth}}, \bibinfo {author} {\bibfnamefont
  {C.}~\bibnamefont {Bradac}},\ and\ \bibinfo {author} {\bibfnamefont
  {I.}~\bibnamefont {Aharonovich}},\ }\bibfield  {title} {\bibinfo {title}
  {{Identifying carbon as the source of visible single-photon emission from
  hexagonal boron nitride}},\ }\href
  {https://doi.org/10.1038/s41563-020-00850-y} {\bibfield  {journal} {\bibinfo
  {journal} {Nature Materials}\ }\textbf {\bibinfo {volume} {20}},\ \bibinfo
  {pages} {321} (\bibinfo {year} {2021})}\BibitemShut {NoStop}%
\bibitem [{\citenamefont {Stern}\ \emph {et~al.}(2022)\citenamefont {Stern},
  \citenamefont {Gu}, \citenamefont {Jarman}, \citenamefont {Barker},
  \citenamefont {Mendelson}, \citenamefont {Chugh}, \citenamefont {Schott},
  \citenamefont {Tan}, \citenamefont {Sirringhaus}, \citenamefont
  {Aharonovich},\ and\ \citenamefont {Atatüre}}]{10.1038/s41467-022-28169-z}%
  \BibitemOpen
  \bibfield  {author} {\bibinfo {author} {\bibfnamefont {H.~L.}\ \bibnamefont
  {Stern}}, \bibinfo {author} {\bibfnamefont {Q.}~\bibnamefont {Gu}}, \bibinfo
  {author} {\bibfnamefont {J.}~\bibnamefont {Jarman}}, \bibinfo {author}
  {\bibfnamefont {S.~E.}\ \bibnamefont {Barker}}, \bibinfo {author}
  {\bibfnamefont {N.}~\bibnamefont {Mendelson}}, \bibinfo {author}
  {\bibfnamefont {D.}~\bibnamefont {Chugh}}, \bibinfo {author} {\bibfnamefont
  {S.}~\bibnamefont {Schott}}, \bibinfo {author} {\bibfnamefont {H.~H.}\
  \bibnamefont {Tan}}, \bibinfo {author} {\bibfnamefont {H.}~\bibnamefont
  {Sirringhaus}}, \bibinfo {author} {\bibfnamefont {I.}~\bibnamefont
  {Aharonovich}},\ and\ \bibinfo {author} {\bibfnamefont {M.}~\bibnamefont
  {Atatüre}},\ }\bibfield  {title} {\bibinfo {title} {{Room-temperature
  optically detected magnetic resonance of single defects in hexagonal boron
  nitride}},\ }\href {https://doi.org/10.1038/s41467-022-28169-z} {\bibfield
  {journal} {\bibinfo  {journal} {Nature Communications}\ }\textbf {\bibinfo
  {volume} {13}},\ \bibinfo {pages} {618} (\bibinfo {year} {2022})}\BibitemShut
  {NoStop}%
\bibitem [{\citenamefont {Li}\ \emph {et~al.}(2021)\citenamefont {Li},
  \citenamefont {Wang}, \citenamefont {Yan}, \citenamefont {Zhou},
  \citenamefont {Wang}, \citenamefont {Liu}, \citenamefont {Guo}, \citenamefont
  {Zhou}, \citenamefont {Gali}, \citenamefont {Liu}, \citenamefont {Wang},
  \citenamefont {Sun}, \citenamefont {Guo}, \citenamefont {Tang}, \citenamefont
  {Li}, \citenamefont {You}, \citenamefont {Xu}, \citenamefont {Li},\ and\
  \citenamefont {Guo}}]{10.1093/nsr/nwab122}%
  \BibitemOpen
  \bibfield  {author} {\bibinfo {author} {\bibfnamefont {Q.}~\bibnamefont
  {Li}}, \bibinfo {author} {\bibfnamefont {J.-F.}\ \bibnamefont {Wang}},
  \bibinfo {author} {\bibfnamefont {F.-F.}\ \bibnamefont {Yan}}, \bibinfo
  {author} {\bibfnamefont {J.-Y.}\ \bibnamefont {Zhou}}, \bibinfo {author}
  {\bibfnamefont {H.-F.}\ \bibnamefont {Wang}}, \bibinfo {author}
  {\bibfnamefont {H.}~\bibnamefont {Liu}}, \bibinfo {author} {\bibfnamefont
  {L.-P.}\ \bibnamefont {Guo}}, \bibinfo {author} {\bibfnamefont
  {X.}~\bibnamefont {Zhou}}, \bibinfo {author} {\bibfnamefont {A.}~\bibnamefont
  {Gali}}, \bibinfo {author} {\bibfnamefont {Z.-H.}\ \bibnamefont {Liu}},
  \bibinfo {author} {\bibfnamefont {Z.-Q.}\ \bibnamefont {Wang}}, \bibinfo
  {author} {\bibfnamefont {K.}~\bibnamefont {Sun}}, \bibinfo {author}
  {\bibfnamefont {G.-P.}\ \bibnamefont {Guo}}, \bibinfo {author} {\bibfnamefont
  {J.-S.}\ \bibnamefont {Tang}}, \bibinfo {author} {\bibfnamefont
  {H.}~\bibnamefont {Li}}, \bibinfo {author} {\bibfnamefont {L.-X.}\
  \bibnamefont {You}}, \bibinfo {author} {\bibfnamefont {J.-S.}\ \bibnamefont
  {Xu}}, \bibinfo {author} {\bibfnamefont {C.-F.}\ \bibnamefont {Li}},\ and\
  \bibinfo {author} {\bibfnamefont {G.-C.}\ \bibnamefont {Guo}},\ }\bibfield
  {title} {\bibinfo {title} {{Room temperature coherent manipulation of
  single-spin qubits in silicon carbide with a high readout contrast}},\ }\href
  {https://doi.org/10.1093/nsr/nwab122} {\bibfield  {journal} {\bibinfo
  {journal} {National Science Review}\ }\textbf {\bibinfo {volume} {9}},\
  \bibinfo {pages} {nwab122} (\bibinfo {year} {2021})}\BibitemShut {NoStop}%
\bibitem [{\citenamefont {Foglszinger}\ \emph {et~al.}(2022)\citenamefont
  {Foglszinger}, \citenamefont {Denisenko}, \citenamefont {Kornher},
  \citenamefont {Schreck}, \citenamefont {Knolle}, \citenamefont {Yavkin},
  \citenamefont {Kolesov},\ and\ \citenamefont
  {Wrachtrup}}]{10.1038/s41534-022-00566-8}%
  \BibitemOpen
  \bibfield  {author} {\bibinfo {author} {\bibfnamefont {J.}~\bibnamefont
  {Foglszinger}}, \bibinfo {author} {\bibfnamefont {A.}~\bibnamefont
  {Denisenko}}, \bibinfo {author} {\bibfnamefont {T.}~\bibnamefont {Kornher}},
  \bibinfo {author} {\bibfnamefont {M.}~\bibnamefont {Schreck}}, \bibinfo
  {author} {\bibfnamefont {W.}~\bibnamefont {Knolle}}, \bibinfo {author}
  {\bibfnamefont {B.}~\bibnamefont {Yavkin}}, \bibinfo {author} {\bibfnamefont
  {R.}~\bibnamefont {Kolesov}},\ and\ \bibinfo {author} {\bibfnamefont
  {J.}~\bibnamefont {Wrachtrup}},\ }\bibfield  {title} {\bibinfo {title} {{TR12
  centers in diamond as a room temperature atomic scale vector magnetometer}},\
  }\href {https://doi.org/10.1038/s41534-022-00566-8} {\bibfield  {journal}
  {\bibinfo  {journal} {npj Quantum Information}\ }\textbf {\bibinfo {volume}
  {8}},\ \bibinfo {pages} {65} (\bibinfo {year} {2022})}\BibitemShut {NoStop}%
\end{thebibliography}%
\end{document}